\documentclass[aps,pre,reprint,superscriptaddress,longbibliography]{revtex4-2}

\usepackage[T1]{fontenc}
\usepackage[utf8]{inputenc}
\usepackage{amsmath,amssymb,amsfonts,bm,mathtools}
\usepackage{graphicx}
\usepackage{physics}
\usepackage{hyperref}
\usepackage{xcolor}

\hypersetup{
    colorlinks=true,
    linkcolor=blue,
    citecolor=blue,
    urlcolor=blue
}

\begin{document}

\title{Bosonic Working Media in a Frustrated Rhombi Chain: Otto and Stirling Cycles from Flat Bands, Caging, and Flux Control}

\author{Francisco J. Pe{\~n}a}
\email{francisco.pena@uss.cl}
\affiliation{
Facultad de Ingenier\'ia, Universidad San Sebasti\'an,
Lago Panguipulli 1390, Puerto Montt, Chile}

\author{Rafael García-Zamora}
\affiliation{Departamento de Física, Universidad Técnica Federico Santa María, Casilla 110-V, Valparaíso, Chile}

\author{Gabriele De Chiara}
\affiliation{Department of Physics, Universitat Autònoma de Barcelona, 08193 Bellaterra (Barcelona), Spain}

\author{Jorge Flores}
\affiliation{Departamento de Física, Universidad Técnica Federico Santa María, Casilla 110-V, Valparaíso, Chile}

\author{Santiago Henríquez}
\affiliation{Departamento de Física, Universidad Técnica Federico Santa María, Casilla 110-V, Valparaíso, Chile}
\affiliation{Instituto de Física, Pontificia Universidad Católica de Valparaíso, Casilla 4950, 2373223 Valparaíso, Chile}

\author{Nathan M. Myers}
\affiliation{Physical and Computational Sciences,Pacific Northwest National Laboratory, Richland, Washington 99354, USA}

\author{Felipe Barra}
\affiliation{Departamento de Física, Facultad de Ciencias Físicas y Matemáticas, Universidad de Chile, Santiago, Chile}

\author{Patricio Vargas}
\affiliation{Departamento de Física, Universidad Técnica Federico Santa María, Casilla 110-V, Valparaíso, Chile}
\date{\today}

\begin{abstract}
We demonstrate that flat-band engineering provides a direct route to control and optimize the thermodynamic performance of quantum heat engines. We consider noninteracting bosons on a rhombic chain lattice described by a Bose–Hubbard model in the noninteracting limit, where a magnetic flux serves as a tunable parameter that continuously reshapes the single-particle spectrum. By driving the system toward the fully frustrated Aharonov–Bohm caging regime, the band structure transitions from dispersive to completely flat, strongly modifying the thermal occupation of the modes. We show that this flux-induced spectral restructuring has clear and measurable thermodynamic consequences. In particular, the Otto cycle exhibits a significant enhancement of both work output and efficiency when operating near the caging regime. We identify the underlying mechanism as a pronounced suppression of heat released to the cold reservoir, rather than an increase in absorbed heat, revealing that flat-band formation is an effective strategy to increase work extraction. In contrast, the Stirling cycle is governed by entropy changes during isothermal, flux-driven processes, leading to greater work extraction over a broader parameter range but at lower efficiency. These results establish geometric frustration and Aharonov–Bohm caging as thermodynamic resources and show that spectral engineering via synthetic gauge fields offers a viable, experimentally accessible pathway to tailor the performance of bosonic quantum thermal machines.
\end{abstract}

\maketitle
\section{Introduction}

Quantum thermal machines provide a natural framework for exploring how the microscopic properties of a working medium determine its macroscopic thermodynamic performance. In small and mesoscopic systems, quantities such as work output, heat exchange, and efficiency can depend sensitively on the structure of the energy spectrum, the presence of degeneracies, particle statistics, and the nature of external control parameters~\cite{Kosloff2013,Vinjanampathy2016,Bhattacharjee2021,Mukherjee2021,Deffner2019,Esposito2009,Campisi2011,Seifert2012,Goold2016,Quan2007,Kieu2004}. This sensitivity has motivated an extensive search for working substances in which genuinely quantum, many-body, or spectral effects can be harnessed as thermodynamic resources, as illustrated by recent studies of statistics-driven many-body engines, nonlinear bosonic working media, prethermal quantum Otto cycles, superconducting quantum heat engines, and interferometric bosonic cycles~\cite{Koch2023PauliEngine,Chesi2025BosonicEngine,Brollo2025PrethermalEngine,Uusnakki2026SuperconductingEngine,Ferreri2025InterferometricEngine}.

Among the most promising candidates are lattice systems whose band structure can be controlled through geometry, interference, or gauge fields, either natural or synthetic. In such systems, external parameters do not merely shift individual energy levels, but can reorganize the entire spectrum, modify the density of states, and reshape the thermal occupation of low-energy modes. Flat-band systems are particularly appealing in this context, as they naturally host compact localized states, macroscopic degeneracies, and strong spectral compression without the need for disorder~\cite{Leykam2018,Derzhko2015,Flach2014,Sutherland1986}. Beyond these general features, diamond-chain and related frustrated lattices provide paradigmatic realizations in which flat bands coexist with dispersive ones and can be continuously tuned via external parameters~\cite{Ahmed2022AllBandFlat,Marques2023Kaleidoscopes}. Highlighting the thermodynamic potential of spectrally engineered systems, recent work on magic-angle twisted bilayer graphene has shown that flat Landau levels can strongly reshape the operational regimes and thermodynamic performance of Otto, Carnot, and Stirling cycles, reinforcing the broader connection between spectral compression, flat-band thermodynamics, and quantum thermal machines~\cite{Soufy2025FlatBandThermodynamics}.

A paradigmatic mechanism leading to flat bands is Aharonov--Bohm caging, where destructive interference induced by a gauge flux suppresses wave-packet spreading and yields completely dispersionless bands at special frustration points~\cite{Vidal1998,Mosseri2022GapLabeling}. Early theoretical work on the rhombi chain addressed interaction-induced delocalization and the effects of disorder, while related fully frustrated Josephson-junction chains were shown to support the pairing of Cooper pairs~\cite{Vidal2000ABcaging,Vidal2001DisorderABCages,Doucot2002Pairing}. These foundational concepts paved the way for artificial-device realizations, such as the implementation of Josephson-junction rhombi in superconducting nanocircuits designed for protected-qubit operation~\cite{Gladchenko2009ProtectedQubits}.

Building on these early models, this localization phenomenon has been extensively studied in a variety of flat-band systems---ranging from interacting and disordered regimes to few-body dynamics~\cite{Pelegri2019ABOAM,Pelegri2020TwoBosons,Roy2020DiamondDisorder}---and experimentally observed in synthetic platforms such as photonic lattices and engineered waveguide arrays~\cite{Mukherjee2018,Kremer2020PhotonicGauge}. More recently, Aharonov--Bohm caging and related localization phenomena have also been explored in ultracold atoms, topolectrical circuits, and broader superconducting platforms, providing increasing experimental access to flux-controlled flat-band physics~\cite{Li2022UltracoldABCaging,Wang2022TopolectricalIAT,Zhou2023CircuitABBosons,Rosen2025SuperconductingRhombic}. From a spectral perspective, Aharonov--Bohm caging represents an extreme form of band flattening in which transport is fully suppressed while the density of states becomes strongly concentrated.

More generally, synthetic gauge fields provide a powerful route to engineer band structures and topological properties in quantum matter. These techniques have been successfully implemented in ultracold atoms and photonic systems, enabling the realization of paradigmatic models such as the Hofstadter and Harper Hamiltonians~\cite{Aidelsburger2013Gauge,Miyake2013Gauge,Dalibard2011,Goldman2014,Ozawa2019}. In this context, the ability to reshape the energy landscape via external control parameters suggests a natural pathway for engineering thermodynamic functionality.

In this work, we consider a bosonic rhombi-chain (diamond-chain) lattice described by a Bose--Hubbard Hamiltonian. While the underlying lattice-boson framework is standard~\cite{Fisher2000BosonLocalization}, the specific rhombi-chain geometry and its frustration effects were detailed in Ref.~\cite{Cartwright2018}. Here, we specialize the model to the noninteracting limit. In this regime, the system reduces to a quadratic hopping Hamiltonian, fully determined by its single-particle spectrum, providing a particularly transparent setting for isolating the role of geometric frustration and interference effects. When a magnetic or synthetic gauge flux threads each plaquette, the spectrum evolves continuously from a dispersive regime to a fully frustrated point at which all bands become flat and the eigenstates form compact Aharonov--Bohm cages. The magnetic flux therefore acts as a genuine spectral control parameter, allowing one to tune the system between mobile and localized regimes without modifying the lattice connectivity.

The central question we address is whether such flux-induced spectral restructuring can be exploited as a thermodynamic resource in bosonic heat engines. More specifically, we investigate how band-structure deformation affects heat exchange, work extraction, and efficiency in flux-driven Otto and Stirling cycles. Because we operate in the noninteracting limit, the many-body thermodynamics is entirely determined by the exact single-particle spectrum, enabling a direct connection between band-structure properties and thermodynamic performance.

Our results show that the two thermodynamic cycles probe the same flux-controlled spectrum in qualitatively different ways. In the Otto protocol, approaching the Aharonov--Bohm caging regime enhances performance by strongly suppressing the heat released to the cold reservoir, while only moderately affecting the heat absorbed from the hot reservoir. This asymmetry increases both the extracted work and the efficiency. In contrast, the Stirling cycle is governed by entropy variations along isothermal, flux-driven branches. As a result, it produces more work over a broader parameter range but at lower efficiency, reflecting its greater sensitivity to entropy variations during isothermal processes.

These findings demonstrate that geometric frustration and flat-band formation are not merely spectral or transport features; they leave clear thermodynamic fingerprints that can be harnessed to optimize the performance of quantum thermal machines. More broadly, they show that gauge control of lattice spectra provides a powerful route to engineer thermodynamic functionality directly at the level of the energy landscape.

The remainder of the paper is organized as follows. In Sec.~\ref{sec:model} we introduce the rhombi-chain bosonic model and its exact flux-dependent band structure. In Sec.~\ref{sec:thermo} we formulate the equilibrium thermodynamic framework. In Sec.~\ref{sec:caging} we analyze the thermodynamic signatures of Aharonov--Bohm caging. Sections~\ref{sec:otto} and~\ref{sec:stirling} present the flux-driven Otto and Stirling cycles, respectively, and compare their performance across different spectral regimes. In Sec.~\ref{sec:finite_time} we introduce a phenomenological finite-time extension of the Otto cycle and analyze its efficiency at maximum power. In Sec.~\ref{sec:experimental_feasibility} we discuss experimental feasibility and candidate platforms. Finally, in Sec.~\ref{sec:conclu} we summarize our results and outline possible directions for future work.
\section{Rhombi-chain bosonic model}
\label{sec:model}

\begin{figure}[t]
\centering
\includegraphics[width=\linewidth]{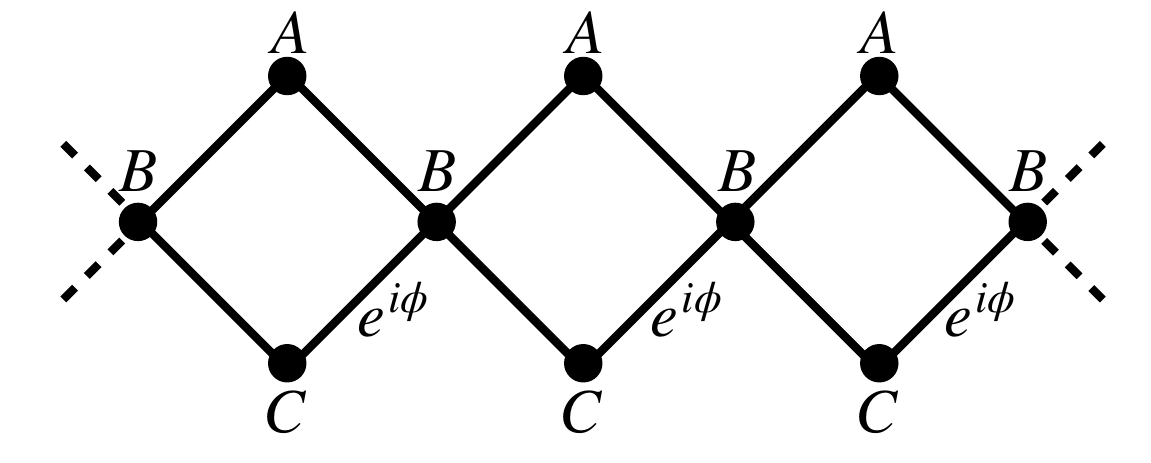}
\caption{
Schematic representation of the rhombi-chain (diamond-chain) lattice used as the working medium. The lattice consists of three sublattice sites per unit cell, labeled $A$, $B$, and $C$, connected in a rhombic geometry. A magnetic flux $\phi$ threads each plaquette, giving rise to phase-dependent hopping amplitudes through the Peierls substitution, as indicated by the factors $e^{i\phi}$. This flux-controlled geometry induces interference effects that can lead to flat-band formation and Aharonov--Bohm caging, which play a central role in the thermodynamic behavior analyzed in this work.
}
\label{fig:rhombi_chain}
\end{figure}

We consider bosonic quasiparticles on a one-dimensional rhombi-chain lattice threaded by a magnetic flux $\phi$ through each plaquette, as illustrated in Fig.~\ref{fig:rhombi_chain}. Each unit cell contains three sites labeled $A$, $B$, and $C$, forming a diamond-like (rhombic) structure that naturally gives rise to geometric frustration and interference effects. The connectivity of the lattice, together with the presence of a flux piercing each plaquette, allows for the emergence of nontrivial band structures that can be continuously tuned by varying $\phi$. Such diamond-chain geometries are paradigmatic realizations of flat-band systems, where geometric frustration and interference effects lead to compact localized states and macroscopic degeneracies \cite{Leykam2018,Derzhko2015,Flach2014}.

The magnetic flux serves as a genuine control parameter that modifies the phase acquired by particles hopping around each plaquette, thereby reshaping the single-particle spectrum without altering the underlying lattice geometry. As a consequence, the system can be driven from a dispersive regime, where the bands exhibit finite curvature and extended states, to the fully frustrated Aharonov--Bohm caging regime \cite{Vidal1998,Vidal2000ABcaging,Mosseri2022GapLabeling}, where destructive interference suppresses transport, and all bands become flat. In this limit, the eigenstates are compactly localized within individual plaquettes, providing a clear manifestation of interference-driven localization, extensively studied in rhombic lattices and related flat-band systems \cite{Pelegri2019ABOAM,Pelegri2020TwoBosons,Roy2020DiamondDisorder}.

Since our primary goal is to investigate the thermodynamic response induced by this flux-controlled spectral structure, we focus on the noninteracting limit of the Bose--Hubbard model ($U=0$). In this regime, the many-body problem reduces to a quadratic hopping Hamiltonian that is entirely determined by the single-particle spectrum. This simplification allows us to establish a direct and transparent connection between band-structure properties and thermodynamic quantities such as internal energy, entropy, and specific heat, which form the basis for the analysis of the quantum thermal cycles considered in this work.

\subsection{Single-particle Hamiltonian}

The Hamiltonian reads
\begin{equation}
\hat H_0 =
-J
\sum_j
\sum_{\ell=0,\pm 1}
\sum_{\alpha,\beta}
T^{(\ell)}_{\alpha\beta}
\,
\hat b^{\dagger}_{j+\ell,\alpha}
\hat b_{j,\beta},
\end{equation}
where $J>0$ is the hopping amplitude, and the indices $\alpha,\beta \in \{A,B,C\}$ label the three sublattice sites within each unit cell. The matrices $T^{(\ell)}$ encode both the connectivity of the rhombi lattice and the Peierls phases associated with the magnetic flux.

A convenient gauge choice for a flux $\phi$ through each rhombus is
\begin{align}
T^{(0)} &=
\begin{pmatrix}
0 & 1 & 0\\
1 & 0 & 1\\
0 & 1 & 0
\end{pmatrix},
\\
T^{(+1)} &=
\begin{pmatrix}
0 & 1 & 0\\
0 & 0 & e^{i\phi}\\
0 & 0 & 0
\end{pmatrix},
\\
T^{(-1)} &= \bigl(T^{(+1)}\bigr)^\dagger .
\end{align}

For a translationally invariant infinite system, it is convenient to
work in momentum space. Introducing the spinor
\begin{equation}
\hat{\bm b}_k=
\begin{pmatrix}
\hat b_{k,A}\\
\hat b_{k,B}\\
\hat b_{k,C}
\end{pmatrix},
\end{equation}
the Hamiltonian becomes
\begin{equation}
\hat H_0 =
\sum_k
\hat{\bm b}_k^\dagger
\,H(k,\phi)\,
\hat{\bm b}_k ,
\end{equation}
with the Bloch Hamiltonian
\begin{equation}
H(k,\phi)=
-J
\begin{pmatrix}
0 & 1+e^{-ik} & 0\\
1+e^{ik} & 0 & 1+e^{i(\phi-k)}\\
0 & 1+e^{-i(\phi-k)} & 0
\end{pmatrix}.
\label{eq:BlochHam}
\end{equation}

\subsection{Exact spectrum and Aharonov--Bohm caging}

Diagonalizing the Bloch Hamiltonian yields three energy branches
labeled by $\tau=0,\pm1$,

\begin{equation}
E_{\tau}(k,\phi)=
2J\tau
\sqrt{1+\cos\!\left(k-\frac{\phi}{2}\right)
\cos\!\left(\frac{\phi}{2}\right)}.
\label{eq:spectrum}
\end{equation}

This spectrum exhibits several remarkable properties. Similar spectral features, including the coexistence of flat and dispersive bands and their evolution under flux control, have been widely analyzed in diamond-chain and related frustrated lattices \cite{Ahmed2022AllBandFlat,Marques2023Kaleidoscopes}.

First, the central band $\tau=0$ is perfectly flat for all values of the flux $\phi$. Second, the upper and lower bands become progressively flatter as the flux approaches $\phi=\pi$. At the fully frustrated point,

\begin{equation}
E_{\tau}(k,\pi)=2J\tau,
\qquad \tau=0,\pm1,
\end{equation}

all three bands become completely flat and independent of momentum.

In this regime the group velocity

\begin{equation}
v_{\tau}(k)=
\frac{1}{\hbar}
\frac{\partial E_{\tau}}{\partial k}
\end{equation}

vanishes identically, and single-particle transport is suppressed by destructive interference. The corresponding eigenstates are compact localized states known as \emph{Aharonov--Bohm cages}.

From the perspective of thermal machines, the magnetic flux therefore acts as a spectral control parameter: by varying $\phi$, one can tune the bosonic working medium between dispersive regimes and fully localized flat-band regimes.

\subsection{Positive excitation spectrum}

The spectrum in Eq.~(\ref{eq:spectrum}) contains both positive and negative branches.
While this is natural from the band-structure perspective, it is convenient to work with a strictly positive excitation spectrum when interpreting the system as a bosonic quasiparticle gas, for instance, in magnonic realizations.

To this end, we introduce a constant energy shift
\begin{equation}
\widetilde E_\tau(k,\phi)
=
E_0 + E_\tau(k,\phi),
\label{eq:shifted_spectrum}
\end{equation}
with $E_0$ chosen such that
$\widetilde E_\tau(k,\phi)>0$ for all $(k,\phi)$.

The required value of $E_0$ can be obtained from the minimum of the original spectrum. From Eq.~(\ref{eq:spectrum}) one finds that the
lowest energy occurs for the branch $\tau=-1$ when the argument of the square root is maximal, yielding
\begin{equation}
E_{\min} = -2J\sqrt{2}.
\end{equation}
Therefore, choosing
\begin{equation}
E_0 = 2J\sqrt{2},
\end{equation}
guarantees that the shifted spectrum remains strictly positive
throughout the Brillouin zone. In the numerical calculations, we use the
equivalent decimal value $E_0 \approx 2.83$ (with $J=1$).

In second quantization, the shifted Hamiltonian reads

\begin{equation}
\widetilde H =
\sum_{k,\tau}
\widetilde E_\tau(k,\phi)
\,b^\dagger_{k\tau}b_{k\tau}
=
H
+
E_0
\sum_{k,\tau}
b^\dagger_{k\tau}b_{k\tau}.
\end{equation}

The additional term corresponds to $E_0 N$, where $N$ is the total number of particles. Therefore, the shift is equivalent to a redefinition of the chemical potential in the grand-canonical ensemble and does not affect the thermodynamic derivatives that determine the caloric response of the
system.

Physically, this procedure amounts to measuring the quasiparticle energies relative to a reference ground state, in close analogy with
standard descriptions of magnon spectra in magnetic systems.

\section{Thermodynamic framework}
\label{sec:thermo}

The thermodynamic properties of the working medium follow directly from the
single-particle spectrum introduced in Sec.~\ref{sec:model}. Throughout this
work, we use the shifted excitation energies
$\widetilde E_\tau(k,\phi)$ defined in
Eq.~(\ref{eq:shifted_spectrum}), which ensure that the bosonic spectrum
remains strictly positive over the entire Brillouin zone and for all values
of the magnetic flux.

Unless otherwise stated, all numerical results presented in
the remainder of this work are expressed in natural units with
$J=k_{\mathrm B}=1$. Consequently, temperatures are measured in units of the
hopping energy $J$, and all energies, heats, and works are given in the same
units. This convention is adopted throughout the thermodynamic analysis and
in all subsequent figures.

We describe the system as a gas of noninteracting bosonic particles in thermal equilibrium within the grand-canonical ensemble, characterized by temperature $T$ and chemical potential $\mu$. Since the excitation spectrum is strictly positive, we set $\mu=0$, as is appropriate for nonconserved bosonic excitations. This choice provides a consistent equilibrium description while avoiding the unphysical divergences that would otherwise arise from the occupation of the lowest-energy mode. In this way, the thermodynamic response is entirely determined by the flux-dependent spectrum and the corresponding Bose occupation factors.

The occupation number of a bosonic mode labeled by $(k,\tau)$ is given by
\begin{equation}
n_B\!\left(\widetilde E_\tau(k,\phi)\right)
=
\frac{1}{e^{\beta \widetilde E_\tau(k,\phi)}-1},
\qquad
\beta=\frac{1}{k_B T},
\label{eq:bose}
\end{equation}
and, for convenience, we define
\begin{equation}
n_{k,\tau}\equiv n_B\!\left(\widetilde E_\tau(k,\phi)\right).
\label{eq:nktau}
\end{equation}

\subsection{Grand potential}

The equilibrium thermodynamics of the system is obtained from the grand potential,
\begin{equation}
\Omega(T,\phi)
=
k_B T
\sum_{k,\tau}
\ln\!\left[
1-e^{-\beta \widetilde E_\tau(k,\phi)}
\right],
\label{eq:grandpotential}
\end{equation}
where $\mu=0$ has already been incorporated, consistently with the description of nonconserved bosonic particles and the positive-definite excitation spectrum.

All relevant thermodynamic observables follow from this quantity through standard equilibrium relations. In particular, the total particle number is
\begin{equation}
N(T,\phi)
=
\sum_{k,\tau} n_{k,\tau},
\label{eq:N}
\end{equation}
while the internal energy is
\begin{equation}
U(T,\phi)
=
\sum_{k,\tau}
\widetilde E_\tau(k,\phi)\, n_{k,\tau}.
\label{eq:U}
\end{equation}

Because the system is noninteracting, these quantities are fully determined by the single-particle spectrum and its thermal occupation. This makes the present framework particularly well-suited to identifying how changes in the flux-modified band structure translate directly into equilibrium thermodynamic properties.

\subsection{Entropy and specific heat}

The entropy is obtained from the temperature derivative of the grand potential at fixed flux,
\begin{equation}
S(T,\phi)
=
-\left(
\frac{\partial \Omega}{\partial T}
\right)_{\phi},
\label{eq:Sdef}
\end{equation}
which provides the natural thermodynamic measure of the redistribution of bosonic occupations induced by temperature and magnetic flux.

An equivalent explicit expression is
\begin{equation}
S
=
k_B
\sum_{k,\tau}
\left[
(1+n_{k,\tau})\ln(1+n_{k,\tau})
-
n_{k,\tau}\ln n_{k,\tau}
\right].
\label{eq:Sexplicit}
\end{equation}
This form is particularly useful for the numerical evaluation of the entropy landscape in the $(T,\phi)$ plane, which later serves to identify adiabatic trajectories and to characterize the thermodynamic response near the Aharonov--Bohm caging regime.

The specific heat at constant flux is defined as
\begin{equation}
C_\phi
=
\left(
\frac{\partial U}{\partial T}
\right)_\phi.
\label{eq:Cphi}
\end{equation}
This quantity provides direct information about how thermal energy is redistributed across the spectrum as the temperature changes. In the present context, it also offers a useful diagnostic of the flux-induced restructuring of the energy bands, since modifications of the spectral density and band flatness leave characteristic signatures in the caloric response.

\subsection{Flux-controlled thermodynamics}

In this system, the magnetic flux enters the thermodynamics exclusively through the single-particle spectrum. Varying $\phi$ therefore deforms the entire set of energy levels and, as a consequence, modifies the Bose occupation distribution of the working medium.

From a thermodynamic perspective, the magnetic flux acts as an external control parameter that reshapes the energy landscape without altering the lattice geometry itself. This spectral control is the key ingredient for implementing flux-driven thermodynamic cycles, since variations of $\phi$ modify the internal energy, entropy, and specific heat through their dependence on the band structure. In this sense, the flux plays a role analogous to a generalized work parameter: by changing $\phi$, one drives the system between different spectral configurations, thereby enabling work extraction through controlled deformation of the bosonic particle spectrum.

This observation establishes the conceptual basis for the analysis developed in the following sections. Once the equilibrium functions $U(T,\phi)$, $S(T,\phi)$, and $C_\phi(T,\phi)$ are known, one can systematically explore how the approach to the Aharonov--Bohm caging regime reorganizes the thermodynamic response and how this spectral restructuring can be exploited in flux-driven Otto and Stirling cycles.


\section{Thermodynamic fingerprints of Aharonov--Bohm caging}
\label{sec:caging}

Throughout the following sections, thermodynamic quantities are obtained by summing over the Brillouin zone, with the momentum $k$ spanning the interval $[-\pi,\pi]$. In the continuum limit, this corresponds to integrating over the full Brillouin zone.

The magnetic flux $\phi$ acts as an external control parameter that modifies the energy spectrum but does not affect the domain of the momentum. Owing to its phase-like nature, the spectrum is periodic in $\phi$ with period $2\pi$, so it is sufficient to restrict the analysis to the fundamental interval $\phi \in [-\pi,\pi]$.

It is therefore important to distinguish between the roles of $k$ and $\phi$: the former is a dynamical variable over which thermodynamic quantities are computed, while the latter parametrizes the deformation of the spectrum without altering the integration domain.

Finally, all temperatures are expressed in units of the hopping amplitude $J$.

\subsection{Internal-energy signatures of Aharonov--Bohm caging}

We analyze the thermodynamic response of the rhombi-chain spectrum as a function of temperature $T$ and magnetic flux $\phi$. Particular attention is given to the fully frustrated point $\phi=\pi$, where Aharonov--Bohm caging occurs and the single-particle band structure becomes strongly reorganized.

To isolate the flux-induced contribution, we define the internal-energy difference
\begin{equation}
\Delta U(T,\phi)=U(T,\phi)-U(T,0),
\end{equation}
which measures the deviation from the zero-flux thermodynamic behavior.

\begin{figure}[t]
\centering
\includegraphics[width=\linewidth]{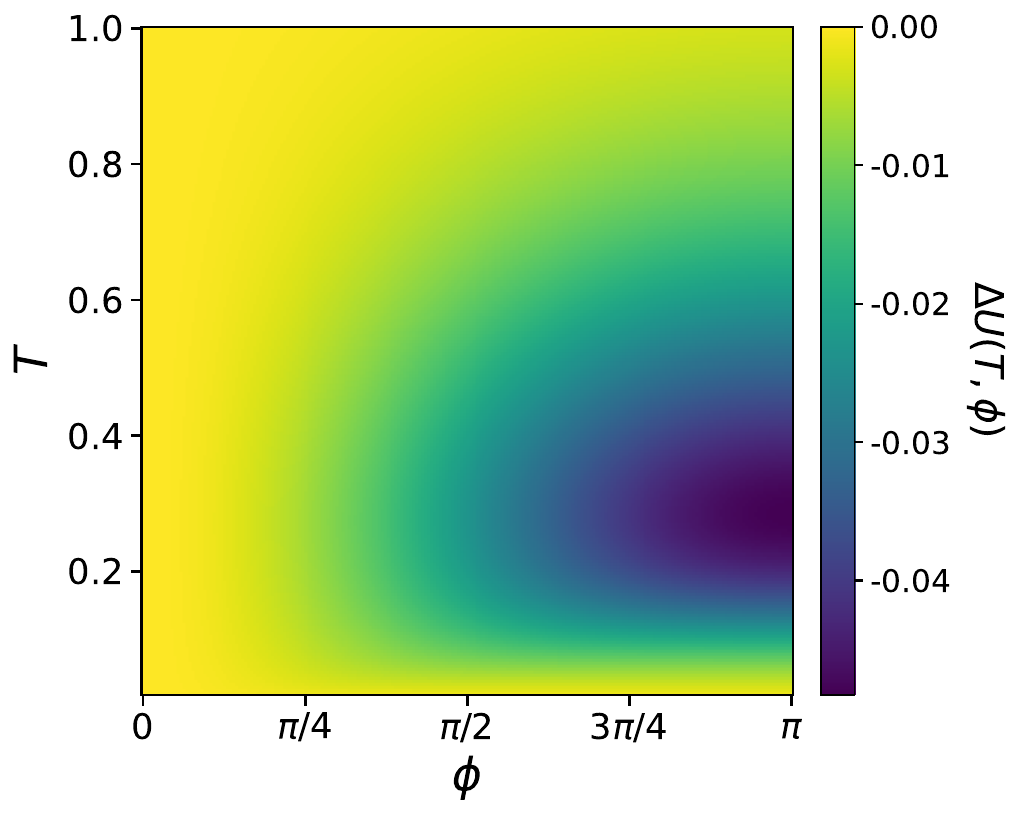}
\caption{
Internal-energy difference
$\Delta U(T,\phi)=U(T,\phi)-U(T,0)$
in the low-temperature regime $0<T/J\leq1$, with
$J=k_{\mathrm B}=1$.
A pronounced negative response develops as the magnetic flux approaches
$\phi=\pi$, where $\Delta U(T,\phi)$ reaches its minimum. This behavior
provides an equilibrium thermodynamic signature of the spectral
reorganization associated with Aharonov--Bohm caging.}
\label{fig:heatmap_lowT}
\end{figure}

The global structure of $\Delta U(T,\phi)$ in the low-temperature regime is shown in Fig.~\ref{fig:heatmap_lowT}. The map exhibits a smooth dependence on $\phi$, with a well-defined minimum emerging near $\phi=\pi$. This minimum becomes more pronounced within a finite temperature window, revealing a clear thermodynamic signature of the caging regime.

\begin{figure}[t]
\centering
\includegraphics[width=\linewidth]{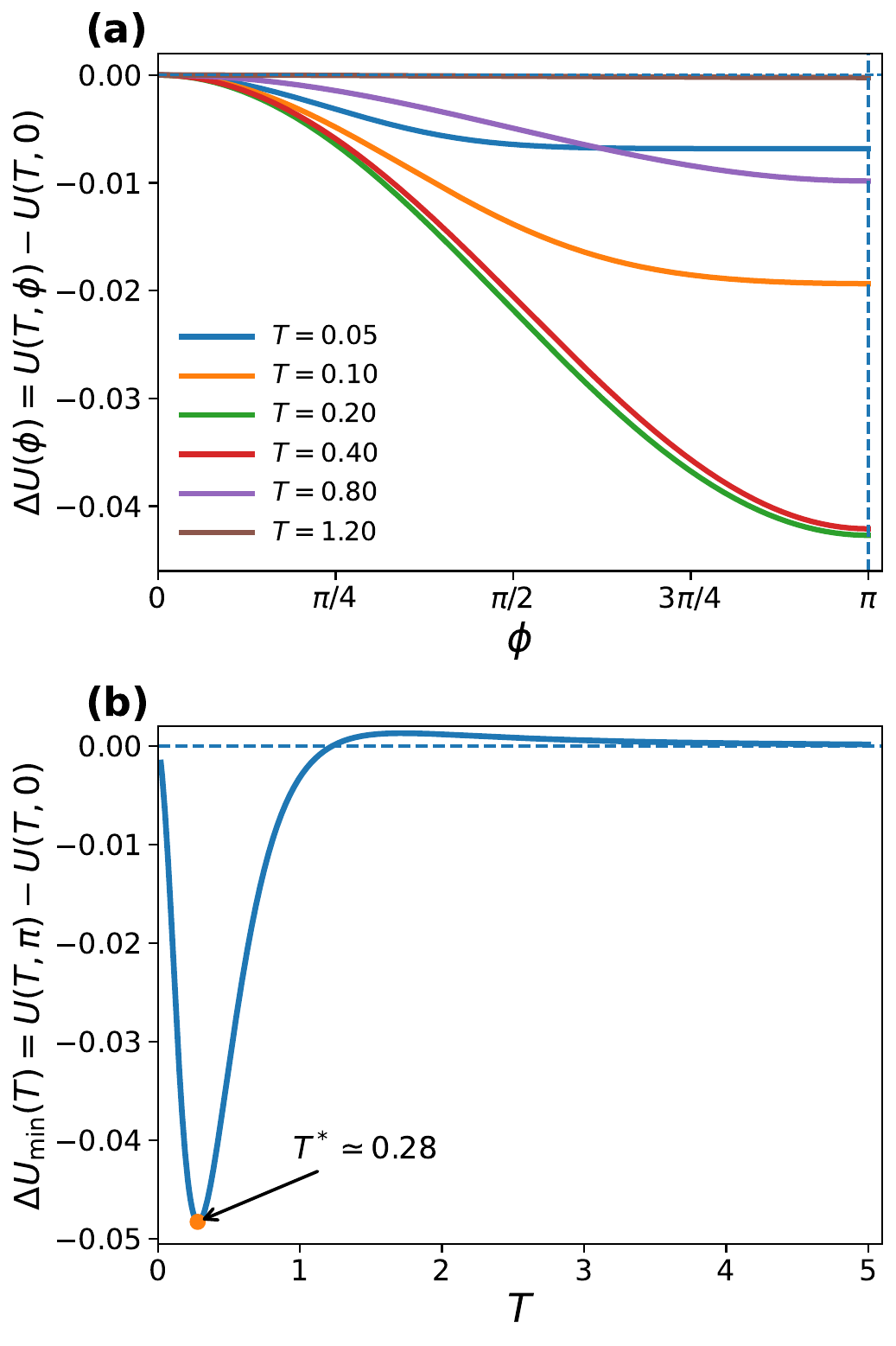}
\caption{
Flux- and temperature-resolved internal-energy response associated with
Aharonov--Bohm caging.
Panel (a): $\Delta U(T,\phi)$ as a function of flux for several fixed
temperatures. For all temperatures shown, the internal-energy difference
decreases as the system approaches the fully frustrated point and reaches
its minimum at $\phi=\pi$. Thus, within the selected parameter range, the
caged spectrum has a lower equilibrium internal energy than the
corresponding zero-flux spectrum.
Panel (b): temperature dependence of
$\Delta U_{\min}(T)=\Delta U(T,\pi)$.
The magnitude of the energy reduction is nonmonotonic and is largest at the
intermediate crossover temperature $T^\ast/J\simeq0.28$. This temperature
marks the regime in which the flux-induced spectral reorganization has its
strongest integrated effect on the equilibrium internal energy.}
\label{fig:cuts}
\end{figure}

This behavior originates from the modification of the single-particle spectrum at the fully frustrated point, where the bands become nearly flat and compact localized states emerge. As a result, the density of low-energy states is enhanced, favoring the population of low-energy modes and lowering the internal energy relative to the zero-flux configuration. To resolve this structure more clearly, Fig.~\ref{fig:cuts} shows cuts of $\Delta U(T,\phi)$ as a function of $\phi$ for several temperatures, together with the temperature dependence of the minimum value $\Delta U_{\min}(T)=\Delta U(T,\pi)$. Several key features emerge. The position of the minimum is robustly pinned at $\phi=\pi$ across all temperatures, confirming that the thermodynamic response is directly tied to the caging condition rather than to a thermal shift of the spectrum. The depth of the minimum exhibits a nonmonotonic dependence on temperature. At low temperatures, only the lowest-energy modes are populated, limiting the impact of the spectral restructuring. As temperature increases, a larger portion of the spectrum contributes, and the enhanced density of low-energy states near $\phi=\pi$ leads to a stronger reduction of the internal energy. At higher temperatures, the occupation becomes broadly distributed, and the differences between flux configurations are progressively washed out by thermal averaging. Consequently, the magnitude of $\Delta U_{\min}(T)$ decreases. Accordingly, $T^\ast$ identifies the intermediate thermal window in which
the flux-induced internal-energy response reaches its largest magnitude.
The physical origin of this crossover is discussed below.


Figures~\ref{fig:heatmap_lowT} and~\ref{fig:cuts} demonstrate that
Aharonov--Bohm caging leaves a robust imprint on the equilibrium
thermodynamics of the bosonic working medium. The effect is not monotonic
in temperature: it is weak at very low temperature, becomes maximal within
an intermediate thermal window, and is progressively reduced at higher
temperature. This behavior establishes a direct connection between
flux-induced spectral restructuring and the macroscopic internal-energy
response.

\begin{figure}[t]
\centering
\includegraphics[width=\linewidth]{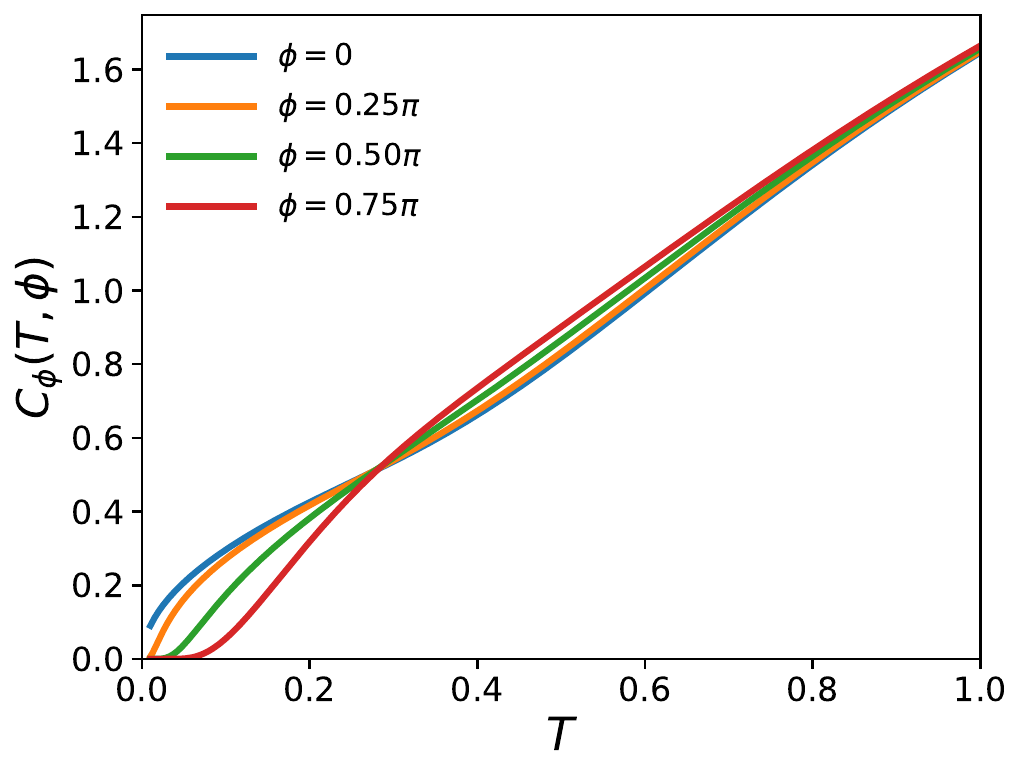}
\caption{
Specific heat at constant flux, $C_\phi(T,\phi)$, as a function of temperature
for several magnetic-flux values, with $J=k_{\mathrm B}=1$.
The curves cross within a narrow intermediate-temperature window close to
the crossover temperature $T^\ast$ at which
$\Delta U_{\min}(T)$ reaches its most negative value.
This crossing is an exact thermodynamic signature of the extremum of
$\Delta U(T,\phi)$ and should be regarded as a diagnostic of the underlying
occupation-driven crossover, rather than as its microscopic origin.}
\label{fig:Cv}
\end{figure}

The physical origin of the crossover temperature $T^\ast$ lies in the
competition between spectral selectivity at low temperature and broad
thermal averaging at high temperature. In the low-temperature regime, only
states close to the bottom of the excitation spectrum are appreciably
occupied. Although the magnetic flux strongly reshapes the spectrum near
the Aharonov--Bohm caging point, the number of thermally active modes remains
limited, so the integrated internal-energy contrast between the caged and
zero-flux configurations is comparatively small.

As the temperature increases, additional momentum and band states become
thermally populated. In this intermediate regime, the bosonic occupation
more efficiently samples the portion of the spectrum that is strongly
reorganized and compressed by the magnetic flux. The resulting spectral
redistribution produces the largest magnitude of
$\Delta U_{\min}(T)$. At still higher temperatures, the occupation spreads
over a broader energy range, and thermal averaging progressively reduces
the contrast between the dispersive and caged spectra. Therefore,
$T^\ast$ identifies the crossover temperature at which these two tendencies
balance in the present model; it is not a universal temperature associated
with flat-band systems.

The specific heat provides a complementary and exact thermodynamic
diagnostic of this crossover. From the definition
$\Delta U(T,\phi)=U(T,\phi)-U(T,0)$, one obtains

\begin{equation}
\frac{\partial}{\partial T}\Delta U(T,\phi)
=
C_\phi(T,\phi)-C_\phi(T,0).
\label{eq:deltaU_specific_heat}
\end{equation}

Consequently, an extremum of $\Delta U(T,\phi)$ occurs whenever
$C_\phi(T,\phi)=C_\phi(T,0)$. In particular, at the caging point,
the condition
$C_\phi(T^\ast,\pi)=C_\phi(T^\ast,0)$
identifies the temperature at which
$\Delta U_{\min}(T)=\Delta U(T,\pi)$
reaches its minimum. Equation~\eqref{eq:deltaU_specific_heat} does not by
itself explain why the crossover occurs at that temperature; rather, it
provides the exact stationarity condition associated with the redistribution
of bosonic thermal occupation described above.

Accordingly, the crossing of the curves in Fig.~\ref{fig:Cv} should be
interpreted as a consistency relation and thermodynamic signature of the
crossover, while its physical origin lies in the competition between the
progressive thermal activation of flux-sensitive states and the eventual
washing out of spectral differences by high-temperature averaging.


\subsection{Thermodynamic geometry and entropy response in the $(T,\phi)$ plane}
\label{subsec:entropy_geometry}

Before constructing explicit thermodynamic cycles, it is useful to analyze the equilibrium thermodynamic structure of the working medium in the $(T,\phi)$ plane. In particular, the entropy $S(T,\phi)$ provides direct information about the adiabatic trajectories of the system, since quasi-static adiabatic transformations are represented by curves of constant entropy.

\begin{figure}[t]
\centering
\includegraphics[width=\linewidth]{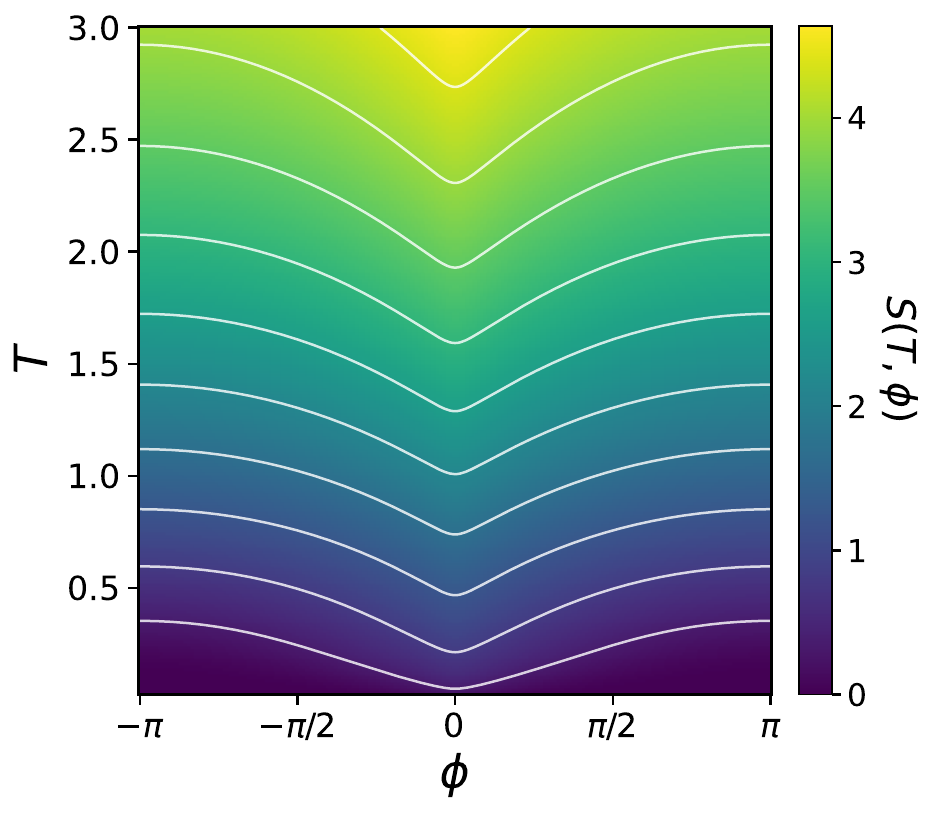}
\caption{
Entropy landscape $S(T,\phi)$ of the rhombi-chain model. The color scale shows the entropy, while the contour lines represent isentropic curves $S=\mathrm{const}$. These curves correspond to the adiabatic trajectories that define the flux-driven branches of an Otto cycle. The map is symmetric under $\phi\to-\phi$ and reveals two mirror-related caloric branches in the $(T,\phi)$ plane.
}
\label{fig:entropy_map}
\end{figure}

Figure~\ref{fig:entropy_map} shows the entropy landscape $S(T,\phi)$ together with several contour lines $S=\mathrm{const}$. These curves represent the isentropic trajectories that define the adiabatic branches of a flux-driven Otto cycle. A key feature of the entropy is its symmetry under the transformation $\phi\to-\phi$. This symmetry follows from the invariance of the single-particle band structure under the combined transformation $(k,\phi)\to(-k,-\phi)$, which ensures that thermodynamic quantities obtained after summation over the Brillouin zone are even functions of $\phi$. As a consequence, the isentropic curves are symmetric around $\phi=0$ and rise smoothly as $|\phi|$ increases.

This structure naturally implies a characteristic caloric phenomenology. Since adiabatic transformations follow the isentropic contours, the slope of these curves determines how the temperature changes when the flux is varied under adiabatic conditions. In the present case, the entropy map indicates opposite caloric responses on the two sides of $\phi=0$: along an isentropic trajectory, increasing $\phi$ on the $\phi<0$ side lowers the temperature, whereas increasing $\phi$ on the $\phi>0$ side raises it. The $(T,\phi)$ plane therefore separates into two mirror-related caloric branches which, due to the $\phi\to-\phi$ symmetry, are thermodynamically equivalent. As a result, flux-driven engine cycles constructed on either side are expected to yield identical performance when defined consistently with the local orientation of the adiabatic branches.

While the absolute entropy $S(T,\phi)$ is a relatively smooth function of flux, this does not imply the absence of thermodynamic signatures associated with frustration or caging. The large temperature-dependent background can obscure the effect of flux modulation when the entropy is plotted directly. For this reason, it is more informative to consider the entropy difference
\begin{equation}
\Delta S(T,\phi)=S(T,\phi)-S(T,0),
\label{eq:DeltaS_def}
\end{equation}
which isolates the entropic response induced by the magnetic flux with respect to the dispersive zero-flux regime.

\begin{figure}[t]
\centering
\includegraphics[width=\linewidth]{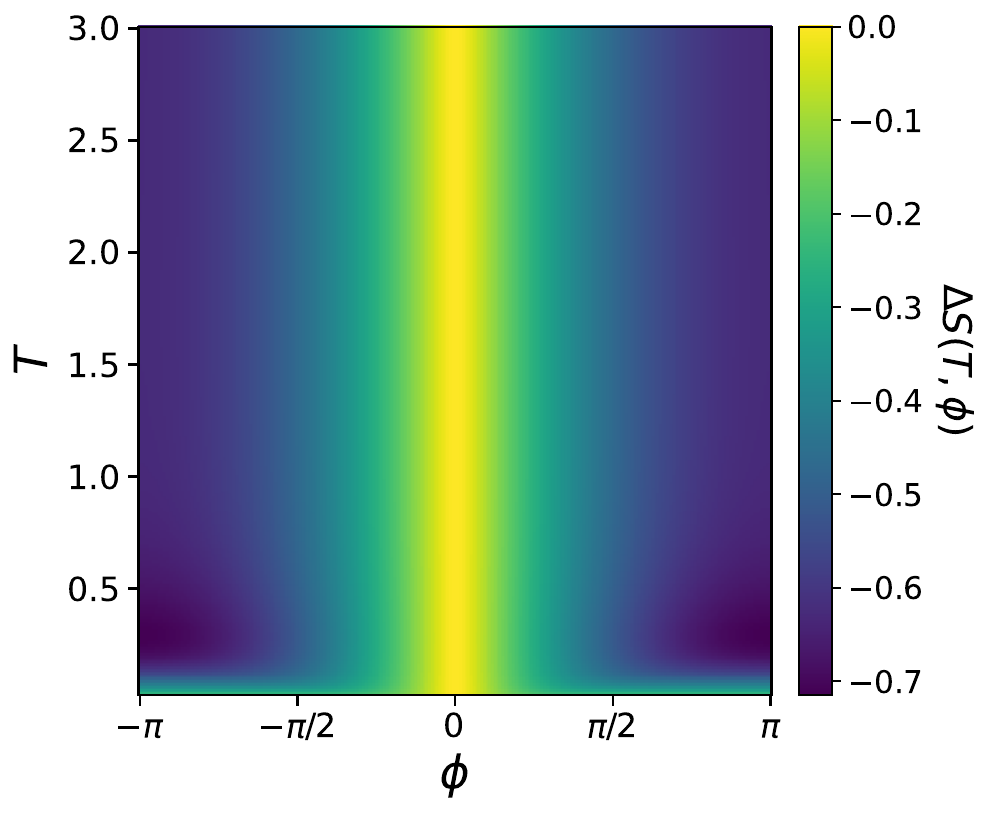}
\caption{
Entropy difference $\Delta S(T,\phi)=S(T,\phi)-S(T,0)$ relative to the zero-flux case. The largest entropy contrast is found close to the fully frustrated points $|\phi|=\pi$, showing that the strongest flux-induced reorganization of the thermal state occurs near the caging regime.
}
\label{fig:DeltaS_map}
\end{figure}

The resulting map, shown in Fig.~\ref{fig:DeltaS_map}, reveals a much clearer structure. The function $\Delta S(T,\phi)$ is even in $\phi$, consistent with the symmetry of the entropy. The entropy variation is negative over most of the diagram, indicating that the application of flux reduces the entropy relative to the $\phi=0$ case. More importantly, the magnitude of this difference is maximized close to the fully frustrated points $|\phi|=\pi$, especially at low and intermediate temperatures. This identifies the caging regime as the region where flux-induced reorganization of the thermal state is strongest.

This result must be interpreted carefully. The map of $\Delta S(T,\phi)$ does not imply that the absolute entropy is maximized at caging. Rather, it shows that the entropic contrast between the dispersive and fully frustrated regimes is largest near $|\phi|=\pi$. In other words, flux modulation produces its strongest effect on the thermal state precisely in the caging regime.

From the perspective of thermal machines, this observation is highly relevant. Since the heat exchanged in isothermal processes is directly proportional to the entropy change, large values of $|\Delta S|$ indicate that the region near $|\phi|=\pi$ is optimal for enhancing caloric effects and thermodynamic performance. Moreover, because adiabatic branches are governed by the geometry of the isentropic contours, the entropy landscape provides a natural framework for designing flux-driven cycles in the $(T,\phi)$ plane.

Taken together, Figs.~\ref{fig:entropy_map} and \ref{fig:DeltaS_map} provide a thermodynamic classification of the model. The zero-flux region corresponds to a dispersive regime, while the vicinity of $|\phi|=\pi$ marks the regime of strongest flux-induced entropic reorganization associated with Aharonov--Bohm caging. In the following sections, this thermodynamic geometry will be used as a guide for constructing explicit engine cycles and comparing the performance of different working regimes.

\section{Flux-driven Otto cycle}
\label{sec:otto}

We now construct a quantum Otto cycle using the bosonic rhombi-chain
working medium introduced in the previous sections.
In this implementation the magnetic flux $\phi$ acts as the external
control parameter that reshapes the single-particle spectrum and
therefore modifies the thermodynamic properties of the system.

The cycle operates between two thermal reservoirs at temperatures
$T_h$ and $T_l$ with $T_h>T_l$.
The magnetic flux controls the spectrum of the working medium,
while the reservoirs control its temperature.

The four corner points of the cycle are defined as

\begin{equation}
A=(T_l,\phi_A), \qquad
B=(T_B,\phi_B),
\end{equation}

\begin{equation}
C=(T_h,\phi_B), \qquad
D=(T_D,\phi_A).
\end{equation}

The cycle is composed of four quasistatic strokes:

\begin{enumerate}

\item[(1)] \textbf{Adiabatic flux increase.}

Starting from state $A=(T_l,\phi_A)$ the system is isolated from the
thermal reservoirs and the magnetic flux is varied from
$\phi_A$ to $\phi_B$.
Since no heat is exchanged during this transformation,
the entropy remains constant,

\begin{equation}
S(T_l,\phi_A)=S(T_B,\phi_B).
\end{equation}

This determines the intermediate temperature $T_B$ and the system
reaches state $B=(T_B,\phi_B)$.

\item[(2)] \textbf{Isoflux heating.}

The system is brought into contact with the hot reservoir at
temperature $T_h$ while the magnetic flux remains fixed at $\phi_B$.
The temperature increases from $T_B$ to $T_h$, bringing the system
to state $C=(T_h,\phi_B)$.

\item[(3)] \textbf{Adiabatic flux decrease.}

The system is again isolated from the thermal reservoirs and the
magnetic flux is changed from $\phi_B$ back to $\phi_A$.
The transformation is isentropic,

\begin{equation}
S(T_h,\phi_B)=S(T_D,\phi_A),
\end{equation}

which determines the temperature $T_D$.
The system therefore reaches state $D=(T_D,\phi_A)$.

\item[(4)] \textbf{Isoflux cooling.}

Finally the system is placed in contact with the cold reservoir at
temperature $T_l$ while the magnetic flux remains fixed at $\phi_A$.
The temperature decreases from $T_D$ to $T_l$, closing the cycle.

\end{enumerate}

The internal energy at the four corner points of the cycle is

\begin{equation}
U_A = U(T_l,\phi_A), \qquad
U_B = U(T_B,\phi_B),
\end{equation}

\begin{equation}
U_C = U(T_h,\phi_B), \qquad
U_D = U(T_D,\phi_A).
\end{equation}

The heat absorbed from the hot reservoir during the heating stroke is

\begin{equation}
Q_h = U_C-U_B ,
\end{equation}
while the heat is released to the cold reservoir during the cooling
stroke is

\begin{equation}
Q_l = U_A-U_D .
\end{equation}

The net work produced during the cycle is therefore

\begin{equation}
W_{\rm net} = Q_h + Q_l = Q_h - |Q_l|
\label{Wotto}
\end{equation}

Positive values of $W_{\rm net}$ correspond to engine operation.
The efficiency of the Otto cycle is then defined as

\begin{equation}
\eta_{\mathrm{Otto}}
=
\frac{W}{Q_h} = 1 + \frac{Q_l}{Q_h} = 1 - \frac{|Q_l|}{Q_h},
\label{Ottoefi}
\end{equation}

In this implementation, the magnetic flux modifies the entire bosonic spectrum of the rhombi-chain model. Changing $\phi$ therefore reshapes the bosonic occupation distribution and alters the internal energy of the working medium.

In particular, the magnetic flux controls the dispersion of the energy bands and can even produce completely flat spectra at the
fully frustrated point $\phi=\pi$. Consequently, the thermodynamic response of the system and therefore the performance of the Otto cycle depend strongly on the chosen values of the flux parameters.

In the following sections, we evaluate the quantities $W_{net}$ and $\eta_{\mathrm{Otto}}$ using the thermodynamic functions $U(T,\phi)$ and $S(T,\phi)$ obtained from the bosonic spectrum. This allows us to systematically explore the performance of the cycle
across different spectral regimes of the rhombi-chain model, ranging from dispersive bands to the strongly frustrated regime near $\phi=\pi$.
\subsection{Example of flux-driven Otto-cycle operation}
\label{subsec:otto_example}

\begin{figure}[t]
\centering
\includegraphics[width=\linewidth]{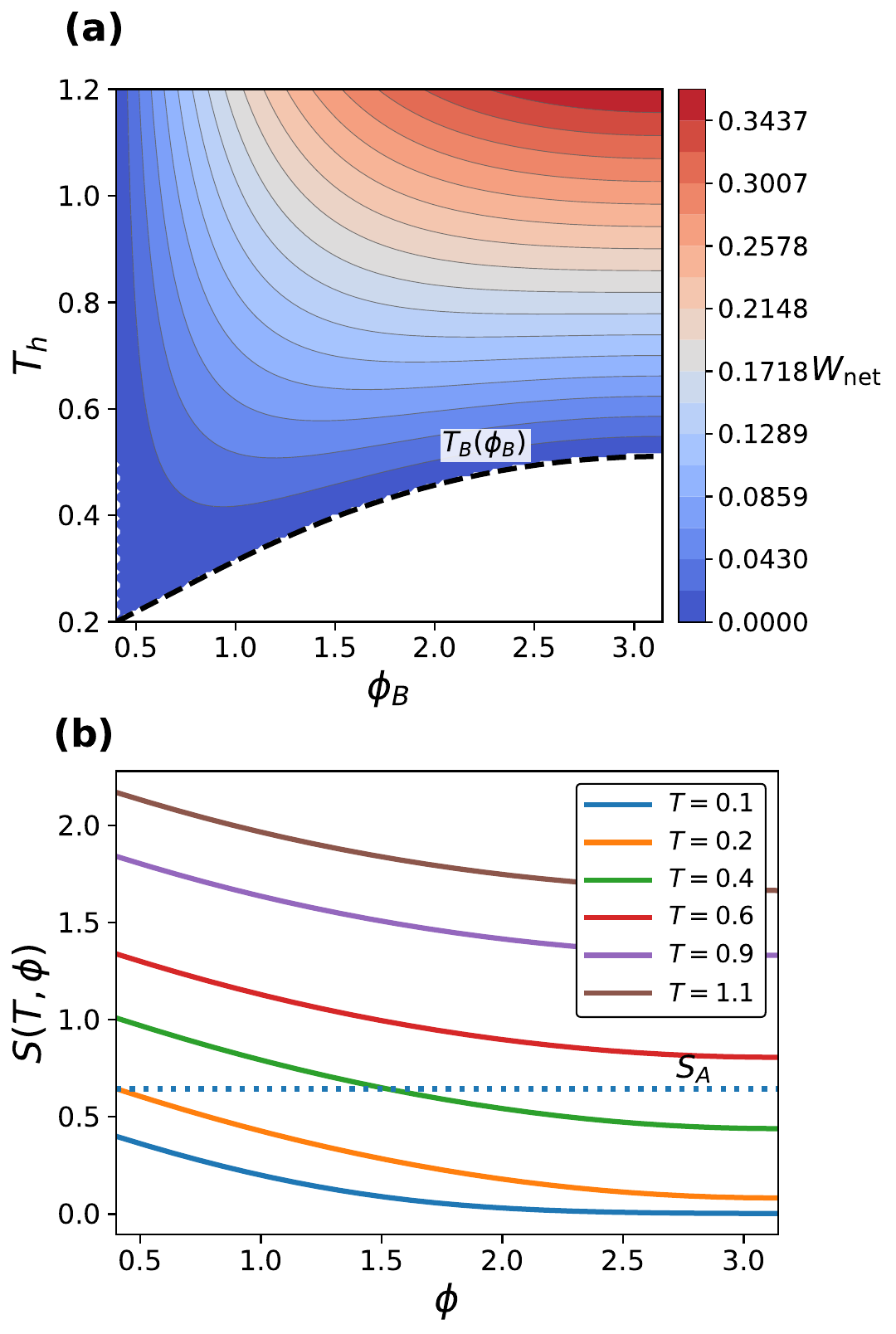}
\caption{
Illustrative execution of the flux-driven Otto cycle for fixed
$\phi_A=0.4$ and $T_l=0.2$.
(a)
Contour map of the net extracted work in the
$(\phi_B,T_h)$ plane.
The dashed curve denotes the isentropic temperature
$T_B(\phi_B)$ determined from the condition
$S(T_B,\phi_B)=S(T_l,\phi_A)$ and therefore marks the lowest hot-reservoir
temperature compatible with the adiabatic compression stroke.
The white region below this curve is thermodynamically inaccessible for
heat-engine operation because the condition
$T_h>T_B(\phi_B)$ is not fulfilled.
The work output increases both with the hot-reservoir temperature and with the degree of spectral compression induced by the magnetic flux, reaching its largest values close to the caging regime.
(b)
Entropy as a function of flux for several fixed temperatures.
The horizontal dotted line indicates the reference entropy
$S_A=S(T_l,\phi_A)$.
The intersections between the entropy curves and this line determine the corresponding isentropic states reached after the first adiabatic stroke. For larger fluxes, the entropy decreases, implying that progressively higher temperatures are required to satisfy the isentropic condition. This behavior explains the upward shift of the boundary $T_B(\phi_B)$ shown in panel (a).}

\label{fig:otto_example}
\end{figure}

To illustrate the execution of the flux-driven Otto cycle, we consider the configuration shown in Fig.~\ref{fig:otto_example}. We fix the initial state of the working medium at $\phi_A=0.4$ and $T_l=0.2$. We then evaluate the net extracted work as a function of the final flux $\phi_B$ and the hot-bath temperature $T_h$.

The upper panel of Fig.~\ref{fig:otto_example} shows that, within the explored parameter window, the net extracted work increases as the flux approaches the strongly frustrated regime. The largest values are observed near the high-flux region, indicating that flux modulation-induced spectral reorganization enhances the thermodynamic response of the cycle.

The lower panel provides a direct thermodynamic interpretation of the forbidden region. Since the first stroke is adiabatic, the entropy remains fixed at its initial value $S_A=S(T_l,\phi_A)$. For each value of $\phi_B$, the corresponding state $B$ is determined by the condition
\begin{equation}
S(T_B,\phi_B)=S(T_l,\phi_A),
\end{equation}
which defines the adiabatic temperature $T_B(\phi_B)$.

If a candidate hot-bath temperature $T_h$ satisfies $S(T_h,\phi_B)<S_A$, then necessarily $T_h<T_B(\phi_B)$, meaning that the hot bath cannot supply heat to the working medium. This explains why the region below the dashed curve in the work map is excluded from engine operation.

As a concrete example, the curve corresponding to $T=0.4$ intersects the dashed entropy level at a well-defined value of $\phi_B$. Beyond this point, the entropy at fixed $T=0.4$ becomes smaller than $S_A$, implying that the corresponding isentropic temperature satisfies $T_B>0.4$. Therefore, if the hot reservoir were set at $T_h=0.4$, the heating stroke would no longer be thermodynamically allowed in that region, and the Otto cycle would cease to operate as a heat engine.


\subsection{Work output along flux and temperature cuts}

To further clarify the structure of the work landscape shown in Fig.~\ref{fig:otto_example}, we examine one-dimensional cuts of the work along the two natural control parameters of the cycle: the final flux $\phi_B$ and the hot-bath temperature $T_h$. Temperatures are expressed in units of $J$.

\begin{figure}[t]
\centering

\includegraphics[width=\linewidth]{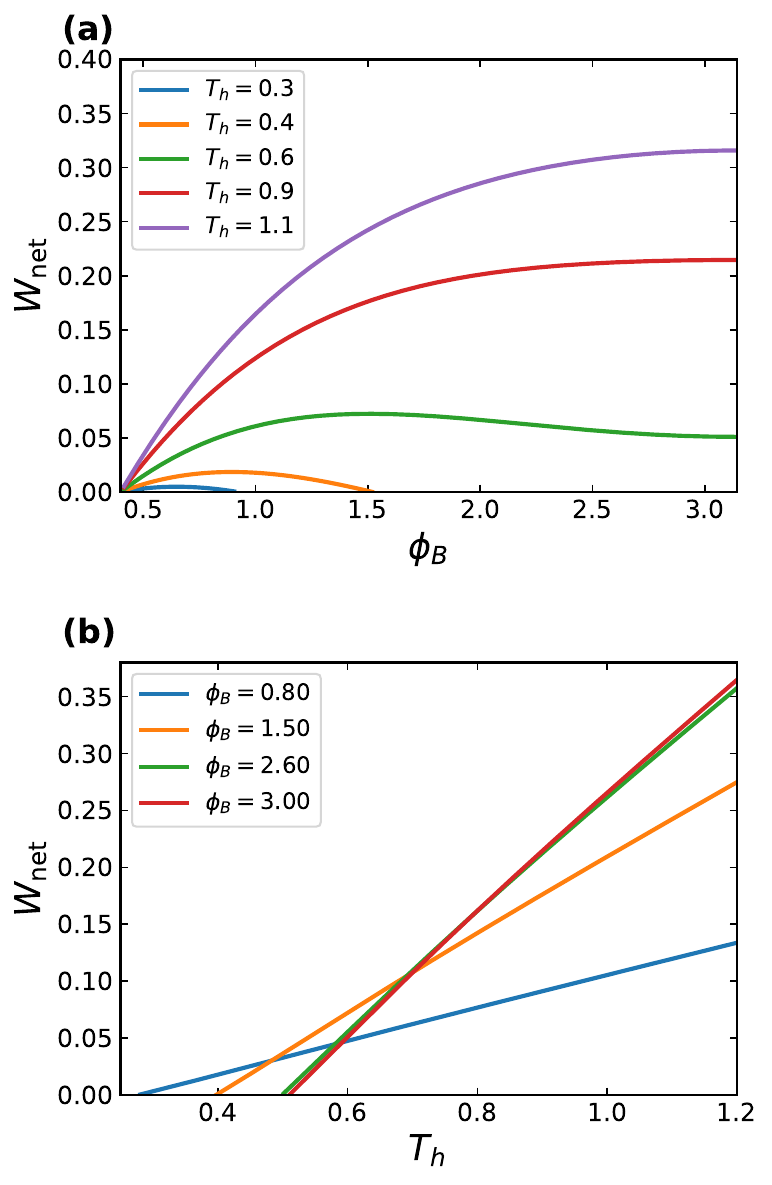}

\caption{Net work extracted from the flux-driven Otto cycle for fixed
$\phi_A=0.4$ and $T_l=0.2$.
Panel (a):}
net work as a function of the final flux $\phi_B$ for several
values of the hot-bath temperature $T_h$.
The work vanishes at $\phi_B=\phi_A$, since no spectral deformation is produced when the flux remains unchanged. As $\phi_B$ increases, the extracted work initially grows because the flux-driven spectral restructuring becomes stronger. For sufficiently low values of $T_h$, the curves terminate before reaching the caging regime because the heating stroke is no longer thermodynamically allowed.
Panel (b):
net work as a function of the hot-bath temperature for several
fixed flux values.
Each curve starts at the threshold temperature
$T_h=T_B(\phi_B)$, above which heat can be absorbed from the hot reservoir. Beyond this threshold, the work increases approximately linearly with $T_h$, while larger values of $\phi_B$ produce steeper slopes and therefore a stronger thermodynamic response.
In both panels, the termination of the curves marks the boundary of the engine-operating region defined by the condition
$T_h>T_B(\phi_B)$.
\label{fig:otto_cuts}
\end{figure}

The upper panel of Fig.~\ref{fig:otto_cuts} shows the dependence of the net work on the final flux $\phi_B$ for several fixed values of the hot-bath temperature. All curves start at $\phi_B=\phi_A$, where the work vanishes because the cycle encloses no thermodynamic area when the flux is unchanged. As $\phi_B$ increases, the extracted work initially grows, reflecting the stronger spectral rearrangement induced by flux modulation. This effect becomes more pronounced for larger values of $T_h$, which provide a greater thermal energy input during the heating stroke.

For sufficiently low hot-bath temperatures, the curves terminate before reaching large values of $\phi_B$. This truncation reflects the thermodynamic constraint that the heating stage requires $T_h>T_B(\phi_B)$. When this condition is not fulfilled, the hot reservoir cannot supply heat, and the cycle cannot operate as a heat engine.

The lower panel of Fig.~\ref{fig:otto_cuts} provides the complementary perspective, showing the work as a function of the hot-bath temperature for several fixed flux values. In this representation, each curve begins at the threshold temperature $T_h=T_B(\phi_B)$, which marks the onset of engine operation. Above this threshold, the work increases approximately linearly with $T_h$ within the explored parameter range.

This near-linear behavior reflects the thermodynamic structure of the cycle. Once the condition $T_h>T_B$ is satisfied, increasing the hot-bath temperature primarily enhances the heat absorbed during the heating stroke, leading to a proportional increase in the extracted work. The slope of the curves increases with $\phi_B$, indicating that larger flux values amplify the thermodynamic response.

Taken together, the two panels highlight the complementary roles of flux and temperature in the engine's operation. The flux controls the spectral rearrangement and determines the accessible thermodynamic states, while the hot-bath temperature regulates the thermal energy injected into the cycle. Their interplay ultimately defines the region of optimal work extraction in the $(\phi_B,T_h)$ plane.
\subsection{Heat-flow asymmetry and work enhancement near caging}

To understand the origin of the work enhancement observed in the previous section, we analyze separately the flux dependence of the heat exchanged with the hot and cold reservoirs, which together determine the net work according to Eq.~\eqref{Wotto}.

\begin{figure}[t]
\centering

\includegraphics[width=\linewidth]{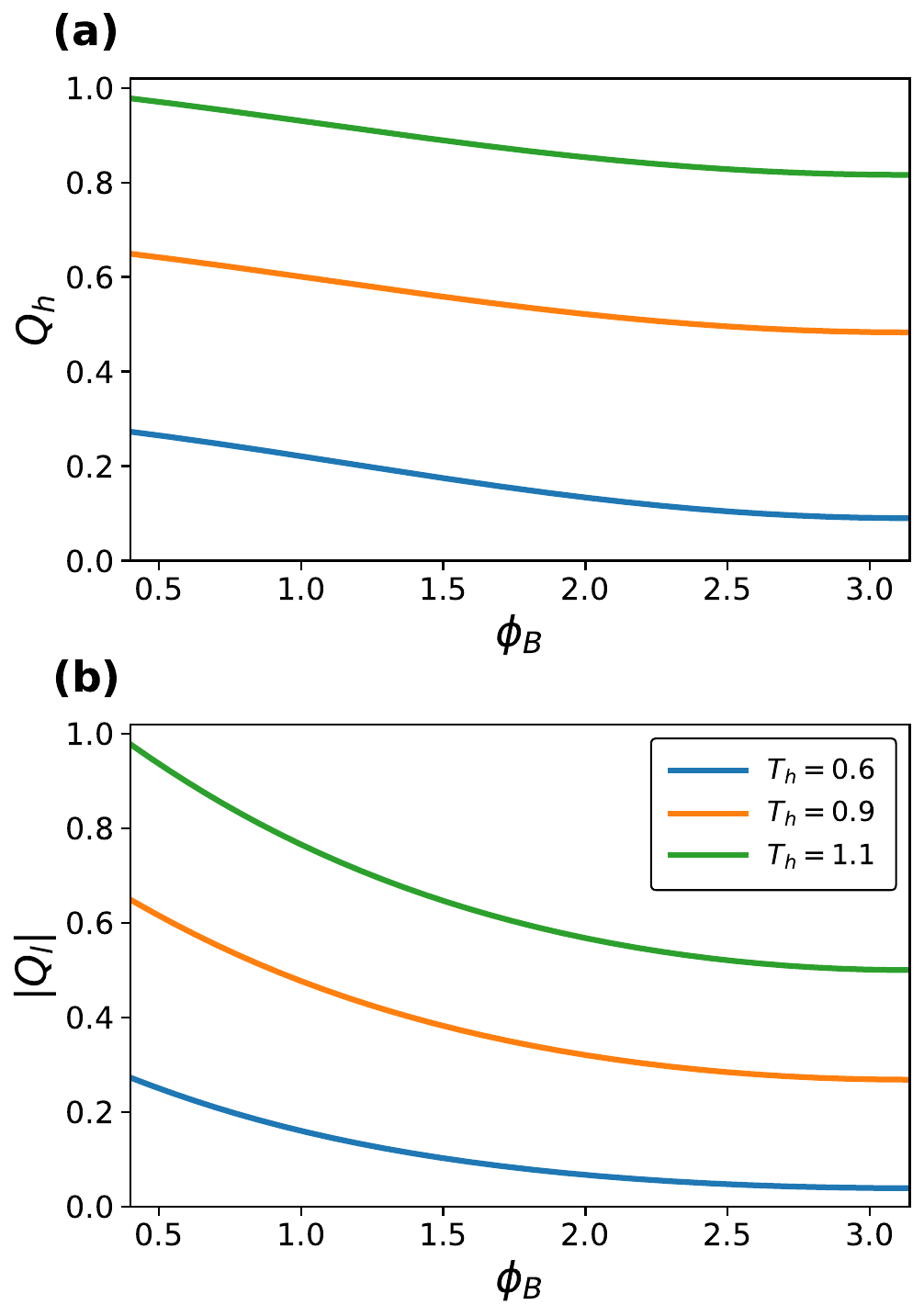}

\caption{
Heat exchanged with the reservoirs as a function of the final flux
$\phi_B$ for fixed $\phi_A=0.4$ and $T_l=0.2$.
Panel (a):}
heat absorbed from the hot reservoir, $Q_h$.
Panel (b):
magnitude of the heat released to the cold reservoir, $|Q_l|$.
The same color coding for the hot-bath temperatures is used in both panels, with the legend displayed only in panel (b).
While $Q_h$ decreases only moderately as $\phi_B$ approaches the
frustrated regime, $|Q_l|$ is strongly suppressed near
$\phi_B\sim\pi$.
Consequently, the increase in
$W_{\rm net}=Q_h-|Q_l|$ is governed primarily by the reduction of the
cold-bath heat loss rather than by an enhancement of the absorbed heat.
This heat-flow asymmetry provides the physical mechanism behind the
improved Otto-cycle performance close to the Aharonov--Bohm caging
regime.

\label{fig:heatflows}
\end{figure}

Figure~\ref{fig:heatflows} shows the heat absorbed from the hot bath $Q_h$ and the magnitude of the heat released to the cold bath $|Q_l|$ as functions of the final flux $\phi_B$, for several values of the hot-bath temperature $T_h$. The initial point of the cycle is fixed at $\phi_A=0.4$ and $T_l=0.2$. Temperatures are expressed in units of $J$.

The origin of the work enhancement can be directly understood from the distinct flux dependence of $Q_h$ and $|Q_l|$. For fixed $T_l=0.2$ and $\phi_A=0.4$, the heat absorbed from the hot bath decreases only moderately as $\phi_B$ approaches the frustrated regime and tends to saturate near $\phi_B\sim\pi$. In contrast, the magnitude of the heat released to the cold bath is strongly suppressed in the same region.

According to Eq.~\eqref{Wotto}, the net work is given by $W_{\mathrm{net}}=Q_h-|Q_l|$. The increase of $W_{\mathrm{net}}$ is therefore dominated by the reduction of $|Q_l|$, rather than by an enhancement of the heat absorbed from the hot bath. This shows that approaching the Aharonov--Bohm caging regime improves work extraction primarily by suppressing heat flow into the cold reservoir.

Physically, this behavior reflects the strong spectral reorganization that
occurs as the magnetic flux approaches the fully frustrated point. Near
$\phi=\pi$, the rhombi-chain spectrum develops nearly flat bands associated
with Aharonov--Bohm caging. This spectral compression modifies the thermal
occupation of the modes and reduces the internal-energy difference between
the two equilibrium states connected by the cold isoflux stroke. As a
result, the magnitude of the heat released to the cold reservoir,
$|Q_l|$, decreases, thereby increasing the net work output.
\subsection{Efficiency landscape and thermodynamic impact of Aharonov--Bohm caging}

To comprehensively summarize the thermodynamic performance of the cycle,
we analyze its efficiency landscape in the $(T_h,\phi_B)$ plane.
According to Eq.~\eqref{Ottoefi}, the efficiency of the Otto cycle is
governed by the balance between the heat absorbed from the hot reservoir
and the heat released to the cold reservoir.

\begin{figure}[t]
\centering
\includegraphics[width=\linewidth]{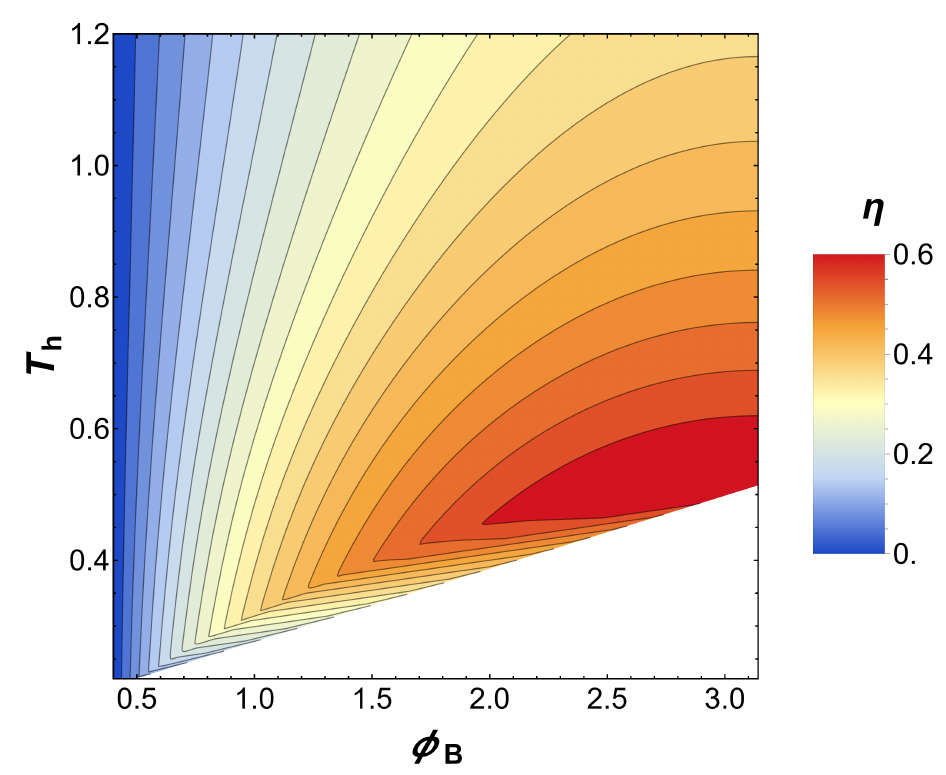}
\caption{
Efficiency landscape $\eta(T_h,\phi_B)$ of the rhombi-chain Otto engine
for $(T_l,\phi_A)=(0.2,0.4)$. The efficiency systematically increases as
the magnetic flux approaches the frustrated Aharonov--Bohm caging regime,
$\phi_B\sim\pi$. The white region corresponds to parameter regimes in
which the cycle can no longer operate as a heat engine.
}
\label{fig:efficiency_map}
\end{figure}

Figure~\ref{fig:efficiency_map} displays the efficiency landscape
$\eta(T_h,\phi_B)$ for the fixed initial configuration
$(T_l,\phi_A)=(0.2,0.4)$. Several key features emerge from the map.
Most notably, the efficiency systematically increases as the magnetic
flux approaches the highly frustrated regime, $\phi_B\sim\pi$. This trend
is consistent with the behavior of the individual heat flows discussed
above. As follows from Eq.~\eqref{Ottoefi}, suppressing the magnitude of
the heat released to the cold reservoir directly enhances the cycle
efficiency.

The physical origin of this performance enhancement lies in the
flux-induced restructuring of the single-particle spectrum. As the
magnetic flux approaches the Aharonov--Bohm caging regime, the progressive
flattening of the energy bands strongly modifies the thermal occupation
of the available modes. This spectral compression reduces the
internal-energy difference between the two equilibrium states connected
by the cold isoflux stroke. Consequently, the working medium releases
significantly less energy to the cold reservoir during the cooling stage,
whereas the heat absorbed from the hot reservoir is affected more
moderately. The enhanced efficiency therefore results from a favorable
asymmetry in heat exchange, driven primarily by suppressed heat rejection
rather than by an increased energy intake from the hot reservoir.

Furthermore, the white region in Fig.~\ref{fig:efficiency_map} has a
direct thermodynamic interpretation. In this region, the temperature
reached after the first adiabatic compression satisfies
$T_B(\phi_B)\geq T_h$. Under this condition, the hot reservoir can no
longer supply heat to the working medium, and the Otto cycle ceases to
operate as a heat engine. These results highlight an
important trade-off: approaching the caging regime improves
heat-to-work conversion through spectral engineering, but simultaneously
reduces the accessible engine-operating domain and the range of viable
hot-bath temperatures.

\section{Flux-driven Stirling cycle}
\label{sec:stirling}

We next analyze a Stirling-type thermodynamic cycle driven by flux modulation. 
The thermodynamic quantities entering the cycle are those introduced in
Sec.~\ref{sec:thermo}, namely the internal energy $U(T,\phi)$
[Eq.~(\ref{eq:U})], the entropy $S(T,\phi)$
[Eq.~(\ref{eq:Sdef})], and the grand potential
$\Omega(T,\phi)$ [Eq.~(\ref{eq:grandpotential})].

To operate the system as a heat engine, the cycle is traversed in the
following order:

\begin{enumerate}
\item[(1)] \textbf{Isoflux heating} at fixed $\phi_H$: $T_l \to T_h$,
\item[(2)] \textbf{Isothermal expansion} at $T_h$: $\phi_H \to \phi_L$,
\item[(3)] \textbf{Isoflux cooling} at fixed $\phi_L$: $T_h \to T_l$,
\item[(4)] \textbf{Isothermal compression} at $T_l$: $\phi_L \to \phi_H$.
\end{enumerate}

Here $\phi_H>\phi_L$. The four thermodynamic states of the cycle are
\begin{equation}
\begin{aligned}
A&=(T_l,\phi_H), \qquad B=(T_h,\phi_H),\\
C&=(T_h,\phi_L), \qquad D=(T_l,\phi_L).
\end{aligned}
\end{equation}

\subsection*{Heat and work along the cycle}

Along the isoflux branches, the flux is held fixed, and therefore no mechanical work is performed,

\begin{equation}
W_{\mathrm{isoflux}} = 0 .
\end{equation}

The exchanged heat is then given by the variation of the internal
energy [Eq.~(\ref{eq:U})],

\begin{equation}
Q_{AB}=U(T_h,\phi_H)-U(T_l,\phi_H),
\end{equation}

\begin{equation}
Q_{CD}=U(T_l,\phi_L)-U(T_h,\phi_L).
\end{equation}

Along the isothermal branches, the exchanged heat follows from the
entropy change [Eq.~(\ref{eq:Sdef})],

\begin{equation}
Q_{BC}=T_h\left[S(T_h,\phi_L)-S(T_h,\phi_H)\right],
\end{equation}

\begin{equation}
Q_{DA}=T_l\left[S(T_l,\phi_H)-S(T_l,\phi_L)\right].
\end{equation}

The work performed along the isothermal branches is determined by the
variation of the grand potential
[Eq.~(\ref{eq:grandpotential})],

\begin{equation}
W_{BC}=-\left[\Omega(T_h,\phi_L)-\Omega(T_h,\phi_H)\right],
\end{equation}

\begin{equation}
W_{DA}=-\left[\Omega(T_l,\phi_H)-\Omega(T_l,\phi_L)\right].
\end{equation}

The total work produced by the cycle is therefore

\begin{equation}
W_{\mathrm{net}} = W_{BC}+W_{DA},
\end{equation}

while the heat absorbed from the hot reservoir is

\begin{equation}
Q_{\mathrm{in}} = Q_{AB}+Q_{BC}.
\end{equation}

The efficiency of the engine is defined as

\begin{equation}
\eta=\frac{W_{\mathrm{net}}}{Q_{\mathrm{in}}},
\end{equation}

for the regime where $W_{\mathrm{net}}>0$ and $Q_{\mathrm{in}}>0$.


\subsection{Comparison between Otto and Stirling cycles}

The Stirling cycle provides a complementary thermodynamic protocol to the Otto engine analyzed in the previous section. While the Otto cycle involves two isochoric heat exchanges and two adiabatic flux transformations, the Stirling cycle replaces the adiabatic branches with isothermal flux-driven processes. As a result, the thermodynamic response is directly controlled by entropy variations induced by flux-dependent spectral restructuring.

\begin{figure}[t]
\centering
\includegraphics[width=\linewidth]{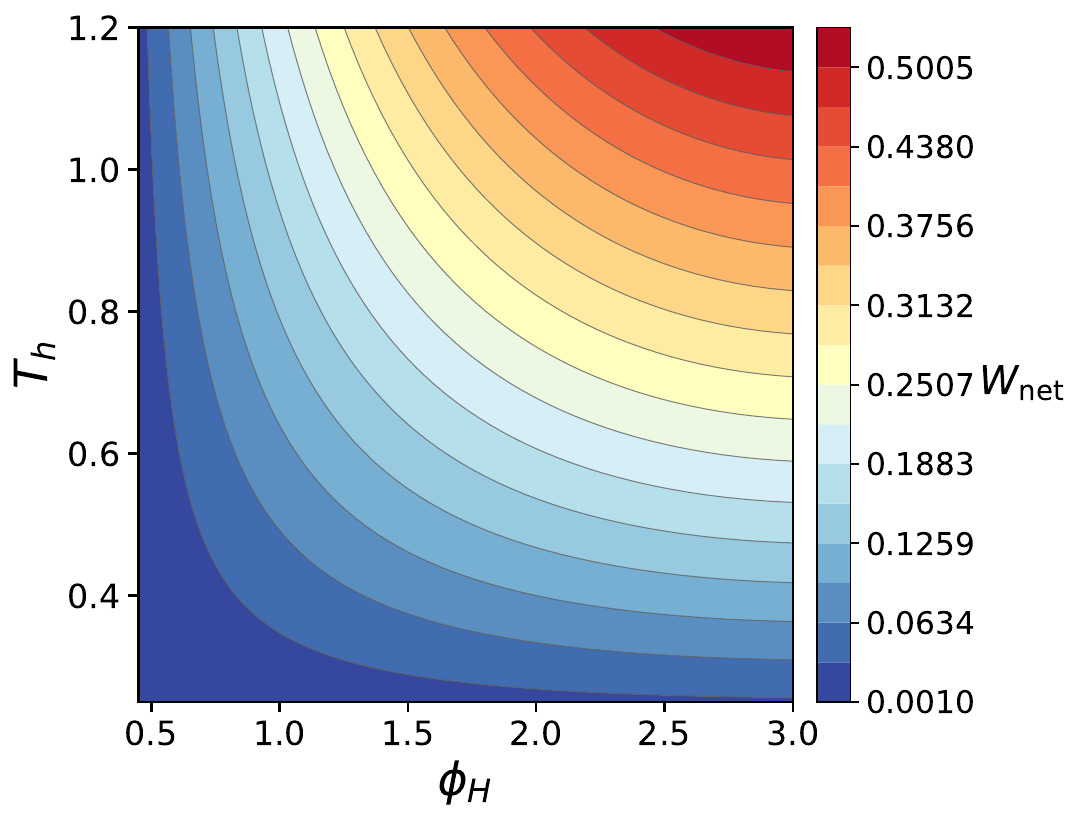}
\caption{
Net work of the Stirling engine in the $(T_h,\phi_H)$ plane for $(T_l,\phi_L)=(0.2,0.4)$. In contrast to the Otto cycle, the Stirling protocol yields positive work over the entire explored region. The work increases smoothly with both the hot temperature and the flux difference $\phi_H-\phi_L$.
}
\label{fig:stirling_work_map}
\end{figure}

Figure~\ref{fig:stirling_work_map} shows the net work of the flux-driven Stirling engine in the $(T_h,\phi_H)$ plane for the same initial conditions used in the Otto analysis, $(T_l,\phi_L)=(0.2,0.4)$, with temperatures expressed in units of $J$. In contrast to the Otto case, where the engine operates only within a restricted region of parameter space, the Stirling protocol produces positive work throughout the explored domain.

This behavior follows from the structure of the cycle. Because the work is generated along the isothermal branches, it is determined by the variation of the grand potential, $W=-\Delta\Omega$. While entropy variations control the heat exchanged along the isothermal processes, the extracted work is not purely entropic in origin. Instead, it reflects the combined effect of entropy and internal-energy changes induced by the flux-dependent deformation of the spectrum. Increasing the flux difference enhances the spectral restructuring experienced during the hot isothermal branch, modifying both the thermal occupation and the energy distribution of the modes. This leads to greater variation in the grand potential and, consequently, to higher extracted work. In addition, increasing $T_h$ amplifies the thermal population of the modes, further enhancing both entropy and internal-energy contributions along the transformation.

A direct comparison with the Otto protocol reveals an important difference in thermodynamic performance. For the same range of parameters, the Stirling cycle generally extracts a larger amount of net work than the Otto engine. However, this increased work output does not translate into a higher efficiency. Instead, the Otto cycle typically achieves higher efficiencies within the same region of the $(T_h,\phi)$ plane.

\begin{figure}[t]
\centering
\includegraphics[width=\linewidth]{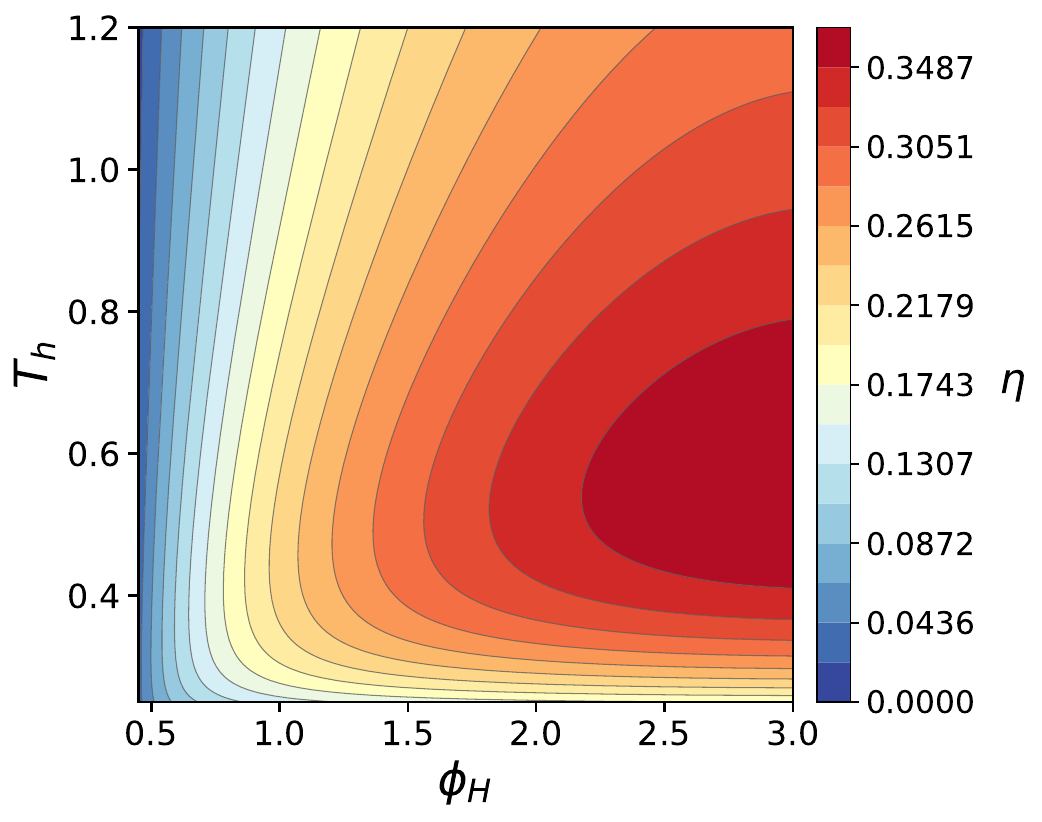}
\caption{
Efficiency landscape of the Stirling engine in the $(T_h,\phi_H)$ plane for $(T_l,\phi_L)=(0.2,0.4)$. The efficiency increases with the flux excursion and reaches its largest values for large $\phi_H$ and moderate temperature differences.
}
\label{fig:stirling_efficiency_map}
\end{figure}

The efficiency landscape of the Stirling engine is shown in Fig.~\ref{fig:stirling_efficiency_map}, which displays a qualitative structure distinct from the Otto case.

The origin of this difference lies in the distinct thermodynamic mechanisms underlying the two cycles. In the Otto protocol, the efficiency enhancement in the high-flux regime arises from the suppression of heat released to the cold bath, thereby increasing both the net work and the efficiency. The Stirling engine, instead, does not rely on such a suppression mechanism. Its performance is governed by entropy variations along the isothermal branches, which increase smoothly with the flux excursion. As a result, the extracted work grows with $\phi_H$, but so does the total heat absorbed from the hot reservoir. Consequently, although the Stirling cycle can extract more work, the ratio $\eta=W_{\mathrm{net}}/Q_{\mathrm{in}}$ remains smaller than in the Otto cycle over the same parameter range.

These results highlight a fundamental distinction between the two protocols. The Otto cycle is highly sensitive to the high-flux regime and exhibits a pronounced thermodynamic response due to reduced heat release. The Stirling cycle, instead, probes the spectrum's global entropy response and therefore displays a smoother thermodynamic landscape dominated by entropy-driven work extraction. Taken together, the two cycles provide complementary thermodynamic diagnostics of flux-induced spectral restructuring in the rhombi-chain system.

\section{Endoreversible finite-time extension}
\label{sec:finite_time}

In the previous sections we demonstrated the thermodynamic advantages of
the rhombi-chain working medium in the quasistatic limit. However, practical heat engines necessarily operate over finite times, where both efficiency and
output power become relevant performance indicators. To address this
point, we introduce a minimal finite-time extension of the Otto cycle using the framework of endoreversibility. Endoreversible thermodynamics is a phenomenological framework that assumes the working medium remains in local equilibrium, but exchanges energy irreversibly with the thermal baths during finite-time heating and cooling strokes that do not necessarily allow the system to fully thermalize to the bath temperatures. This approach to finite-time heat engine analysis was introduced by Curzon and
Ahlborn~\cite{Curzon1975} where they applied it to determine the efficiency at maximum power of the Carnot cycle, now known as the Curzon--Ahlborn efficiency,
\begin{equation}
    \eta_{\rm CA}=1-\sqrt{T_{\rm l}/T_{\rm h}}.
\end{equation}

The endoreversible approach was subsequently formalized and expanded~\cite{Hoffmann1997} and has been applied to many different thermodynamic cycles~\cite{Leff1987}. Despite its apparent similarity to the Carnot efficiency, $\eta_{\rm CA}$ is not a universal upper bound for the efficiency at maximum power \cite{Esposito2010, Deffner2018, Smith2020}. Nevertheless, it provides a useful performance benchmark.

In recent years, endoreversible analysis has
been successfully extended to a broad class of quantum heat engines,
including harmonic oscillators, trapped ions, quantum dots, multilayer
graphene, and plasma-based Stirling
cycles~\cite{Wang2017,Deffner2018,Smith2020,Bouton2021,
Pena2023,Myers2023Graphene,Behrendt2025}. Rather than developing a
complete microscopic nonequilibrium theory, irreversibility is
introduced exclusively through the finite thermalization times, while the
compression and expansion strokes are assumed to remain ideal,
entropy-preserving, and of finite total duration $t_{\rm ad}$. The
present construction is therefore intended as a phenomenological
benchmark that provides a first estimate of finite-time performance
without introducing additional microscopic assumptions.

The equilibrium thermodynamic functions $U(T,\phi)$ and $S(T,\phi)$
derived in the previous sections are retained throughout the cycle.
During each isochoric branch, the effective temperature of the working
medium is assumed to relax exponentially toward the corresponding bath
temperature according to

\begin{equation}
T_{\rm out}
=
T_{\rm bath}
+
\left(
T_{\rm in}
-
T_{\rm bath}
\right)
e^{-t/\tau},
\label{eq:endoreversible}
\end{equation}

where $\tau$ denotes a phenomenological relaxation time and $t$ is the
thermal contact time. The temperature at each corner of the cycle can be solved for using Eq.~\ref{eq:endoreversible} for the isochoric strokes and the constant-entropy conditions $S(T_A,\phi_A)=S(T_B,\phi_B)$ and $S(T_C,\phi_B)=S(T_D,\phi_A)$ for the adiabatic strokes. The work per cycle is evaluated from the same
equilibrium energy differences employed in the quasistatic analysis.
Assuming symmetric thermal contact durations,
$t_{\rm h}=t_{\rm l}=t$, the corresponding output power becomes

\begin{equation}
P
=
\frac{W}{2t+t_{\rm ad}}.
\label{eq:power}
\end{equation}


\begin{figure}[t]
\centering
\includegraphics[width=\columnwidth]{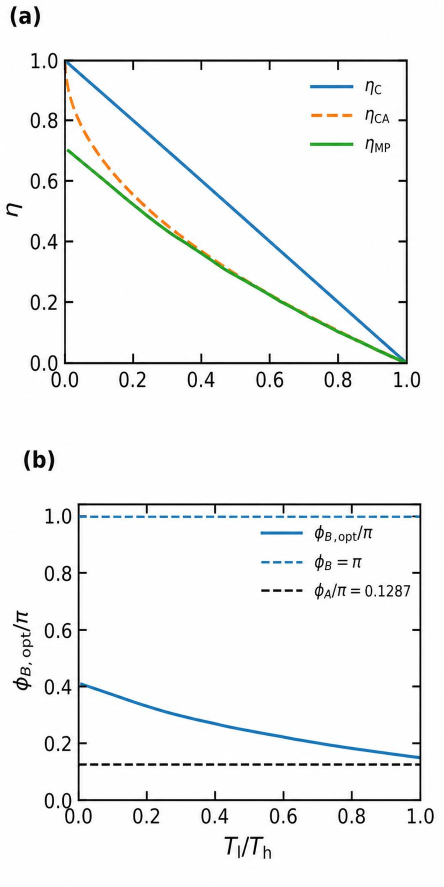}
\caption{
Finite-time performance of the bosonic Otto engine for
$T_{\rm h}/J=0.19$, $\phi_A=0.4$, $t_{\rm h}/\tau=t_{\rm l}/\tau = 1.068$ and
$t_{\rm ad}/\tau=0.4$. 
(a) Carnot efficiency $\eta_{\rm C}$, Curzon--Ahlborn benchmark
$\eta_{\rm CA}$, and efficiency at maximum power $\eta_{\rm MP}$ as
functions of the temperature ratio $T_{\rm l}/T_{\rm h}$.
(b) Corresponding power-optimizing compression flux
$\phi_{B,\rm opt}/\pi$. The blue dashed line marks the complete-caging value
$\phi_B=\pi$, while the black dotted horizontal line indicates the reference level
$\phi_A/\pi = 0.127$.}
\label{fig:EMP}
\end{figure}


In order to determine the optimal power output, an exploratory screening of the thermal parameter space $(T_{\rm h}/J,T_{\rm l}/T_{\rm h})$ was performed. At each set of parameter values, the complete finite-time cycle was solved and the output power
was independently maximized over the admissible compression-flux interval $\phi_A<\phi_B\leq\pi$, retaining only operating points in the engine regime satisfying
$Q_h>0$, $Q_l<0$, and $W>0$. The efficiency was subsequently evaluated at the
optimal value of the compression flux,
\begin{equation}
\eta_{\rm MP}
=
\eta(\phi_{B,\rm opt}),
\quad
\phi_{B,\rm opt}
=
\arg\max_{\phi_A<\phi_B\leq\pi}
P(\phi_B).
\label{eq:eta_MP_flux_optimization}
\end{equation}

In Fig.~\ref{fig:EMP}(a) we plot the efficiency at maximum power for a representative choice of $T_{\rm h}/J=0.19$, selected because it lies within the low-temperature region where the optimized efficiency most closely approaches the Curzon--Ahlborn benchmark while preserving a well-defined heat-engine operating domain. We see that the efficiency at maximum power closely tracks $\eta_{\rm CA}$ for large values of the bath temperature ratio, but falls below it at smaller values of the ratio.

It is important to emphasize that
$\phi_{B,\rm opt}$ is obtained by maximizing the output power and not
the efficiency itself. Each point of Fig.~\ref{fig:EMP}(a) consequently corresponds to a different, independently optimized finite-time cycle evaluated at a fixed compression flux. Correspondingly, Fig.~\ref{fig:EMP}(b) displays the temperature dependence of $\phi_{B,\rm opt}$. Along the selected thermal trajectory, $\phi_{B,\rm opt}$ decreases as $T_{\rm l}/T_{\rm h}$ increases and remains systematically below the complete-caging value $\pi$. Note that for all temperature values $\phi_{B,\rm opt}$ satisfies the imposed condition
$\phi_{B,\rm opt}>\phi_A=0.4$, necessary to ensure the cycle remains in the engine regime.

The displacement of the maximum-power point away from complete caging
can be understood from the contraction of the admissible heat-engine
domain. For the larger working medium temperature ranges permitted by the full thermalization in the
quasistatic analysis, the thermal bias is sufficiently large for the
engine region to extend up to $\phi_B=\pi$, and the output power
increases toward complete caging. Along the low-temperature cut
considered here, however, increasing $T_{\rm l}/T_{\rm h}$
progressively reduces the thermal bias. At sufficiently large
compression flux, the temperature reached after the entropy-preserving
compression stroke may no longer permit heat absorption from the hot
reservoir. Such points fail the heat-engine conditions
$Q_h>0$, $Q_l<0$, and $W>0$ and are consequently excluded from the
admissible operating branch. The finite-time optimum therefore results
from the interplay between flux-induced spectral compression,
incomplete thermalization, the available thermal bias, and the
thermodynamic constraints defining engine operation, rather than from
the caging condition alone.

The present finite-time extension should be interpreted as an initial
phenomenological benchmark rather than a full microscopic nonequilibrium treatment. It neglects
flux-dependent relaxation rates, quantum friction, and explicit
open-system dynamics, all of which may quantitatively modify the
optimal operating conditions. Nevertheless, phenomenological
endoreversible descriptions have proven useful for capturing finite-time
performance across a broad range of quantum working substances
\cite{Wang2017,Deffner2018,Smith2020,Bouton2021,
Pena2023,Myers2023Graphene,Behrendt2025}.



\section{Experimental feasibility and candidate platforms}
\label{sec:experimental_feasibility}

The implementation of the flux-driven thermodynamic cycles proposed in this work relies on a set of physical ingredients that are both conceptually simple and experimentally accessible within current state-of-the-art platforms. In particular, the key requirements are: (i) a rhombi-chain (diamond-chain) lattice geometry supporting multiple sublattice degrees of freedom, (ii) bosonic excitations that can be effectively described within a single-particle framework, and (iii) a tunable gauge flux that continuously reshapes the band structure and drives the system across different spectral regimes, including the fully frustrated Aharonov--Bohm caging point.

A central advantage of the present framework is that, in the noninteracting limit considered here, the full thermodynamic response of the working medium is entirely determined by the single-particle spectrum. In this regime, quantities such as the internal energy, entropy, and specific heat are obtained directly from the flux-dependent dispersion relations and their associated occupation statistics. As a consequence, the proposed thermodynamic cycles do not rely on interaction-induced effects, collective many-body correlations, or fine-tuned coupling mechanisms. Instead, the entire mechanism is governed by spectral engineering through externally controlled phases. This considerably reduces the experimental complexity and shifts the focus toward platforms in which the band structure can be accurately designed, controlled, and probed.

A particularly natural realization of these conditions is provided by synthetic lattice systems with engineered gauge fields. Photonic waveguide arrays, in particular, have emerged as a powerful platform for implementing tight-binding models with tailored geometries and complex hopping amplitudes. In such systems, Aharonov--Bohm caging has been experimentally observed in rhombic lattices, where destructive interference induced by synthetic flux leads to a complete suppression of transport at specific frustration points \cite{Mukherjee2018}. The effective magnetic flux is implemented through controlled phase accumulation along the lattice links, enabling continuous tuning of the band structure from dispersive regimes to completely flat bands. Moreover, recent advances in photonic systems enable the realization of tunable gauge fields and topological band structures, providing a flexible and highly controllable environment for exploring flux-dependent spectral properties \cite{Kremer2020PhotonicGauge}. Beyond these pioneering works, photonic and circuit-based platforms have demonstrated increasingly versatile implementations of flat-band physics and localization phenomena, including topolectrical circuits and superconducting architectures where Aharonov--Bohm caging and related localization effects can be engineered and directly probed \cite{Wang2022TopolectricalIAT,Zhou2023CircuitABBosons,Rosen2025SuperconductingRhombic}.

Ultracold atoms in optical lattices offer a complementary and equally promising route for realizing the ingredients required by our proposal. In these systems, artificial gauge fields can be engineered through laser-assisted tunneling schemes, which enable precise control of Peierls phases and allow for the implementation of lattice Hamiltonians with synthetic magnetic flux \cite{Aidelsburger2013Gauge,Miyake2013Gauge}. Such techniques have already been used to realize paradigmatic models with nontrivial band topology and flux-dependent spectra. More recently, ultracold-atom experiments have directly explored Aharonov--Bohm caging and localization phenomena in controllable lattice geometries, providing clear evidence of flux-induced suppression of transport and tunable spectral flattening \cite{Li2022UltracoldABCaging}. The high degree of control over lattice geometry, tunneling amplitudes, and external parameters makes cold-atom platforms particularly well suited for probing thermodynamic quantities derived from engineered band structures. In addition, the possibility of controlling temperature and coupling to reservoirs in a highly tunable manner provides a natural setting for exploring thermodynamic protocols inspired by quantum heat engines.

From a broader perspective, the framework developed in this work can also be interpreted as describing bosonic quasiparticle excitations in condensed-matter systems. In this context, the introduction of a positive-definite excitation spectrum corresponds to the standard description of bosonic modes defined with respect to a reference ground state, as is commonly done in magnonic or phononic systems. While the direct realization of a full thermodynamic cycle in such systems would require a careful treatment of thermalization processes and controlled coupling to external reservoirs, this viewpoint emphasizes that the underlying mechanism—namely, the control of thermodynamic response through spectral deformation—is not restricted to a specific physical platform. Instead, it represents a general principle that applies across a wide range of bosonic systems in which the excitation spectrum can be engineered or externally tuned.

Importantly, the experimental validation of the present results does not require the immediate realization of a complete quantum heat engine. The central prediction of this work is that the approach to the Aharonov--Bohm caging regime induces a strong reorganization of the system's equilibrium thermodynamic properties. In particular, the suppression of band dispersion near the fully frustrated point leads to a redistribution of thermal occupation toward low-energy modes, which in turn produces measurable signatures in the internal energy and entropy. Within the Otto protocol analyzed here, this effect manifests itself as a strong suppression of the heat released to the cold reservoir, ultimately enhancing both the net work output and the efficiency of the cycle.

A realistic experimental strategy would therefore consist of probing the flux-dependent energy spectrum and reconstructing the corresponding thermal occupation of the modes. From this information, thermodynamic quantities such as the internal energy and entropy can be inferred and compared across different flux regimes. In particular, the contrast between the dispersive regime and the fully frustrated point provides a direct way to identify the thermodynamic fingerprints of Aharonov--Bohm caging. The observation of a flux-induced suppression of energy exchange with the cold reservoir, together with an enhanced thermodynamic contrast in intermediate-temperature windows, would constitute a clear experimental signature of the mechanism described in this work.

In this context, it is worth emphasizing that the most relevant observables are not necessarily transport quantities, but rather equilibrium or quasi-equilibrium properties derived from the spectral structure. This feature further broadens the range of possible experimental implementations by relaxing the need for precise control of currents or nonequilibrium steady states. Instead, measurements of occupation distributions, spectral densities, or effective temperatures can provide indirect yet robust access to the thermodynamic behavior predicted by the theory.

Overall, the combination of rhombic lattice geometries, tunable synthetic gauge fields, and bosonic excitation spectra places the flux-driven thermodynamic cycles investigated here within reach of current experimental capabilities. These platforms offer a realistic and versatile pathway to explore how geometric frustration, flat-band formation, and spectral control can be harnessed as genuine thermodynamic resources in quantum thermal machines. The results presented in this work thus open the door to a new class of experimentally accessible systems in which thermodynamic performance can be engineered through purely geometric and interference-based mechanisms.
\section{Conclusions}
\label{sec:conclu}

We demonstrate that the interplay between magnetic flux and geometric frustration in a bosonic rhombi-chain provides a powerful lever for optimizing quantum thermal machines. By tuning the system toward the Aharonov-Bohm (AB) caging regime ($\phi = \pi$), the working medium undergoes a transition from dispersive transport to complete localization. This spectral restructuring redistributes the thermal occupation of the modes and produces its strongest internal-energy response within an intermediate-temperature window.

In the context of quantum thermal cycles, AB caging acts as an effective control mechanism for heat flow in the Otto protocol: performance enhancement is driven by a pronounced suppression of heat released to the cold reservoir, rather than an increase in heat absorbed from the hot bath. In contrast, the Stirling cycle operates through entropy modulation along isothermal branches, yielding higher work output at the expense of lower efficiency. These results highlight the complementary roles of spectral deformation: heat-flow asymmetry in the Otto cycle and entropic redistribution in the Stirling protocol.

The phenomenological finite-time extension for the Otto cycle demonstrates that the optimal compression flux for maximizing power output is generally displaced from the complete-caging point. These results also demonstrate the
spectral mechanism remains relevant under incomplete thermalization producing deviations of the efficiency at maximum power from the typical Curzon--Ahlborn efficiency in the low-temperature regime. A future fully dynamical treatment developed within an
open-quantum-system framework could explore further finite-time effects including flux-dependent relaxation and
nonadiabatic excitations during finite-speed
driving.

The present analysis is restricted to heat-engine operation. Extending the
flux-controlled Otto and Stirling protocols to refrigerator and heat-pump
regimes, including the corresponding coefficients of performance and
possible reversals of the heat currents, remains an interesting direction
for future work.

Overall, our findings establish synthetic-gauge control as a viable route to engineer thermodynamic functionality directly at the level of the energy landscape. Given that rhombic geometries and tunable gauge fields are already accessible in platforms such as superconducting circuits and ultracold atoms, these flux-controlled effects are within experimental reach. Extending this framework to include interactions and nonequilibrium dynamics will further broaden the design of high-performance lattice-based quantum heat engines.
\vspace{0.2 cm}

\section*{acknowledgments}

F.~J.~P. and P.~V. acknowledge financial support from ANID Fondecyt (Chile) under contracts No.~1230055, 1240582, and 1250173. S.~H. acknowledges support from the ANID (Chile) National Doctoral Fellowship (``Beca de Doctorado Nacional'') No.~21220168, as well as from the ``Programa de Iniciación a la Investigación Científica'' (PIIC) of UTFSM under grant No.~015/2024. P.V. and S.H. also acknowledge support from CEDENNA under grant CIA No.~250002. F.B. acknowledges funding from FONDECYT Regular Grant. 1231210. G. D. C. acknowledges the hospitality of the Universidad de Chile and the Universidad Técnica Federico Santa María, where part of this project has been developed. He acknowledges financial support from the Spanish Agencia Estatal de Investigación project a PID2024-162153NB-I00 and UKRI-EPSRC project UKRI3314. N.~M.~M. acknowledges support from the ``Embedding QC into Many-body Frameworks for Strongly Correlated Molecular and Materials Systems''  project, which is funded by the U.S. Department of Energy, Office of Science, Office of Basic Energy Sciences, the Division of Chemical Sciences, Geosciences, and Biosciences (under FWP 72689).

\bibliographystyle{apsrev4-2}
\bibliography{biblio_APS_revised}

@article{Kosloff2013,
  author  = {Kosloff, Ronnie},
  title   = {Quantum Thermodynamics: A Dynamical Viewpoint},
  journal = {Entropy},
  volume  = {15},
  number  = {6},
  pages   = {2100--2128},
  year    = {2013},
  doi     = {10.3390/e15062100}
}

@article{Vinjanampathy2016,
  author  = {Vinjanampathy, Sai and Anders, Janet},
  title   = {Quantum Thermodynamics},
  journal = {Contemporary Physics},
  volume  = {57},
  number  = {4},
  pages   = {545--579},
  year    = {2016},
  doi     = {10.1080/00107514.2016.1201896}
}

@article{Bhattacharjee2021,
  author  = {Bhattacharjee, Sourav and Dutta, Amit},
  title   = {Quantum thermal machines and batteries},
  journal = {European Physical Journal B},
  volume  = {94},
  pages   = {239},
  year    = {2021},
  doi     = {10.1140/epjb/s10051-021-00235-3}
}

@article{Mukherjee2021,
  author  = {Mukherjee, Victor and Divakaran, Uma},
  title   = {Many-body quantum thermal machines},
  journal = {Journal of Physics: Condensed Matter},
  volume  = {33},
  number  = {45},
  pages   = {454001},
  year    = {2021},
  doi     = {10.1088/1361-648X/ac1b60}
}

@article{Soufy2025FlatBandThermodynamics,
  author    = {Soufy, Hadi Mohammed and Benjamin, Colin},
  title     = {Enhanced performance across {Otto}, {Carnot}, and {Stirling} cycles revealed by flat-band thermodynamics},
  journal   = {Phys. Rev. A},
  volume    = {112},
  pages     = {052215},
  year      = {2025},
  doi       = {10.1103/PhysRevA.112.052215}
}

@book{Deffner2019,
  author    = {Deffner, Sebastian and Campbell, Steve},
  title     = {Quantum Thermodynamics: An Introduction to the Thermodynamics of Quantum Information},
  publisher = {Morgan \& Claypool Publishers},
  address   = {San Rafael, CA},
  year      = {2019},
  doi       = {10.1088/2053-2571/ab21c6}
}

@article{Esposito2009,
  author  = {Esposito, Massimiliano and Harbola, Upendra and Mukamel, Shaul},
  title   = {Nonequilibrium fluctuations, fluctuation theorems, and counting statistics in quantum systems},
  journal = {Reviews of Modern Physics},
  volume  = {81},
  pages   = {1665--1702},
  year    = {2009},
  doi     = {10.1103/RevModPhys.81.1665}
}

@article{Campisi2011,
  author  = {Campisi, Michele and H{\"a}nggi, Peter and Talkner, Peter},
  title   = {Colloquium: Quantum fluctuation relations: Foundations and applications},
  journal = {Reviews of Modern Physics},
  volume  = {83},
  pages   = {771--791},
  year    = {2011},
  doi     = {10.1103/RevModPhys.83.771}
}

@article{Seifert2012,
  author  = {Seifert, Udo},
  title   = {Stochastic thermodynamics, fluctuation theorems and molecular machines},
  journal = {Reports on Progress in Physics},
  volume  = {75},
  number  = {12},
  pages   = {126001},
  year    = {2012},
  doi     = {10.1088/0034-4885/75/12/126001}
}

@article{Goold2016,
  author  = {Goold, John and Huber, Marcus and Riera, Arnau and del Rio, Lidia and Skrzypczyk, Paul},
  title   = {The role of quantum information in thermodynamics---a topical review},
  journal = {Journal of Physics A: Mathematical and Theoretical},
  volume  = {49},
  number  = {14},
  pages   = {143001},
  year    = {2016},
  doi     = {10.1088/1751-8113/49/14/143001}
}

@article{Quan2007,
  author  = {Quan, H. T. and Liu, Yu-xi and Sun, C. P. and Nori, Franco},
  title   = {Quantum thermodynamic cycles and quantum heat engines},
  journal = {Physical Review E},
  volume  = {76},
  pages   = {031105},
  year    = {2007},
  doi     = {10.1103/PhysRevE.76.031105}
}

@article{Kieu2004,
  author  = {Kieu, T. D.},
  title   = {The second law, {Maxwell}'s demon, and work derivable from quantum heat engines},
  journal = {Physical Review Letters},
  volume  = {93},
  pages   = {140403},
  year    = {2004},
  doi     = {10.1103/PhysRevLett.93.140403}
}

@article{Leykam2018,
  author  = {Leykam, Daniel and Andreanov, Alexei and Flach, Sergej},
  title   = {Artificial flat band systems: From lattice models to experiments},
  journal = {Advances in Physics: X},
  volume  = {3},
  number  = {1},
  pages   = {1473052},
  year    = {2018},
  doi     = {10.1080/23746149.2018.1473052}
}

@article{Derzhko2015,
  author  = {Derzhko, Oleg and Richter, Johannes and Maksymenko, Mykola},
  title   = {Strongly correlated flat-band systems: The route from {Heisenberg} spins to {Hubbard} electrons},
  journal = {International Journal of Modern Physics B},
  volume  = {29},
  number  = {27},
  pages   = {1530007},
  year    = {2015},
  doi     = {10.1142/S0217979215300072}
}

@article{Flach2014,
  author  = {Flach, Sergej and Leykam, Daniel and Bodyfelt, J. D. and Matthies, P. and Desyatnikov, A. S.},
  title   = {Detangling flat bands into {Fano} lattices},
  journal = {Europhysics Letters},
  volume  = {105},
  number  = {3},
  pages   = {30001},
  year    = {2014},
  doi     = {10.1209/0295-5075/105/30001}
}

@article{Sutherland1986,
  author  = {Sutherland, Bill},
  title   = {Localization of electronic wave functions due to local topology},
  journal = {Physical Review B},
  volume  = {34},
  pages   = {5208--5211},
  year    = {1986},
  doi     = {10.1103/PhysRevB.34.5208}
}

@article{Ahmed2022AllBandFlat,
  author  = {Ahmed, Aamna and Ramachandran, Ajith and Khaymovich, Ivan M. and Sharma, Auditya},
  title   = {Flat-band-based multifractality in the all-bands-flat diamond chain},
  journal = {Physical Review B},
  volume  = {106},
  pages   = {205119},
  year    = {2022},
  doi     = {10.1103/PhysRevB.106.205119}
}

@article{Marques2023Kaleidoscopes,
  author  = {Marques, A. M. and Pelegr{\'i}, G. and Dias, R. G. and Ahufinger, V. and Mompart, J.},
  title   = {Kaleidoscopes of {Hofstadter} butterflies and {Aharonov--Bohm} caging in rhombus chains},
  journal = {Physical Review Research},
  volume  = {5},
  pages   = {023110},
  year    = {2023},
  doi     = {10.1103/PhysRevResearch.5.023110}
}

@article{Vidal1998,
  author  = {Vidal, Julien and Mosseri, R{\'e}my and Dou{\c{c}}ot, Beno{\^i}t},
  title   = {{Aharonov--Bohm} Cages in Two-Dimensional Structures},
  journal = {Physical Review Letters},
  volume  = {81},
  pages   = {5888--5891},
  year    = {1998},
  doi     = {10.1103/PhysRevLett.81.5888}
}

@article{Vidal2000ABcaging,
  author  = {Vidal, Julien and Dou{\c{c}}ot, Beno{\^i}t and Mosseri, R{\'e}my and Butaud, Patrick},
  title   = {Interaction Induced Delocalization for Two Particles in a Periodic Potential},
  journal = {Physical Review Letters},
  volume  = {85},
  number  = {18},
  pages   = {3906--3909},
  year    = {2000},
  doi     = {10.1103/PhysRevLett.85.3906}
}

@article{Vidal2001DisorderABCages,
  author  = {Vidal, Julien and Butaud, Patrick and Dou{\c{c}}ot, Beno{\^i}t and Mosseri, R{\'e}my},
  title   = {Disorder and interactions in {Aharonov--Bohm} cages},
  journal = {Physical Review B},
  volume  = {64},
  pages   = {155306},
  year    = {2001},
  doi     = {10.1103/PhysRevB.64.155306}
}

@article{Doucot2002Pairing,
  author  = {Dou{\c{c}}ot, Beno{\^i}t and Vidal, Julien},
  title   = {Pairing of {Cooper} Pairs in a Fully Frustrated {Josephson}-Junction Chain},
  journal = {Physical Review Letters},
  volume  = {88},
  pages   = {227005},
  year    = {2002},
  doi     = {10.1103/PhysRevLett.88.227005}
}

@article{Mosseri2022GapLabeling,
  author  = {Mosseri, R{\'e}my and Dou{\c{c}}ot, Beno{\^i}t and Vidal, Julien},
  title   = {{Aharonov--Bohm} cages, flat bands, and gap labeling in rhombus chains},
  journal = {Physical Review B},
  volume  = {106},
  pages   = {155120},
  year    = {2022},
  doi     = {10.1103/PhysRevB.106.155120}
}

@article{Pelegri2019ABOAM,
  author  = {Pelegr{\'i}, G. and Marques, A. M. and Ahufinger, V. and Mompart, J.},
  title   = {Topological edge states and {Aharonov--Bohm} caging with ultracold atoms carrying orbital angular momentum},
  journal = {Physical Review A},
  volume  = {99},
  pages   = {023613},
  year    = {2019},
  doi     = {10.1103/PhysRevA.99.023613}
}

@article{Pelegri2020TwoBosons,
  author  = {Pelegr{\'i}, G. and Marques, A. M. and Ahufinger, V. and Mompart, J.},
  title   = {Interaction-induced topological properties of two bosons in flat-band lattices},
  journal = {Physical Review Research},
  volume  = {2},
  pages   = {033267},
  year    = {2020},
  doi     = {10.1103/PhysRevResearch.2.033267}
}

@article{Roy2020DiamondDisorder,
  author  = {Roy, Nilanjan and Ramachandran, Ajith and Sharma, Auditya},
  title   = {Interplay of disorder and interactions in a flat-band-supporting diamond chain},
  journal = {Physical Review Research},
  volume  = {2},
  pages   = {043395},
  year    = {2020},
  doi     = {10.1103/PhysRevResearch.2.043395}
}

@article{Mukherjee2018,
  author  = {Mukherjee, Sebabrata and Di Liberto, Marco and {\"O}hberg, Patrik and Thomson, Robert R. and Goldman, Nathan},
  title   = {Experimental Observation of {Aharonov--Bohm} Cages in Photonic Lattices},
  journal = {Physical Review Letters},
  volume  = {121},
  number  = {7},
  pages   = {075502},
  year    = {2018},
  doi     = {10.1103/PhysRevLett.121.075502}
}

@article{Kremer2020PhotonicGauge,
  author  = {Kremer, Mark and Petrides, Iacopo and Meyer, Eric and Heinrich, Matthias and Zilberberg, Oded and Szameit, Alexander},
  title   = {A square-root topological insulator with non-quantized indices realized with photonic {Aharonov--Bohm} cages},
  journal = {Nature Communications},
  volume  = {11},
  pages   = {907},
  year    = {2020},
  doi     = {10.1038/s41467-020-14695-2}
}

@article{Gladchenko2009ProtectedQubits,
  author  = {Gladchenko, Sergey and Olaya, David and Dupont-Ferrier, Eva and Dou{\c{c}}ot, Beno{\^i}t and Ioffe, Lev B. and Gershenson, Michael E.},
  title   = {Superconducting nanocircuits for topologically protected qubits},
  journal = {Nature Physics},
  volume  = {5},
  pages   = {48--53},
  year    = {2009},
  doi     = {10.1038/nphys1151}
}

@article{Li2022UltracoldABCaging,
  author  = {Li, Hang and Dong, Zhaoli and Longhi, Stefano and Liang, Qian and Xie, Dizhou and Yan, Bo},
  title   = {{Aharonov--Bohm} Caging and Inverse {Anderson} Transition in Ultracold Atoms},
  journal = {Physical Review Letters},
  volume  = {129},
  pages   = {220403},
  year    = {2022},
  doi     = {10.1103/PhysRevLett.129.220403}
}

@article{Wang2022TopolectricalIAT,
  author  = {Wang, Hailong and Li, Haoran and Liu, Zongping and Lu, Minghui and Chen, Yan-Feng},
  title   = {Observation of inverse {Anderson} transitions in {Aharonov--Bohm} topolectrical circuits},
  journal = {Physical Review B},
  volume  = {106},
  pages   = {104203},
  year    = {2022},
  doi     = {10.1103/PhysRevB.106.104203}
}

@article{Zhou2023CircuitABBosons,
  author  = {Zhou, Xiaobo and Wang, Zhen and Yang, Yue and Xie, Yiming and Guo, Guangcan and Guo, Guoping},
  title   = {Observation of flat-band localization and topological edge states in a superconducting circuit},
  journal = {Physical Review B},
  volume  = {107},
  pages   = {035152},
  year    = {2023},
  doi     = {10.1103/PhysRevB.107.035152}
}

@article{Rosen2025SuperconductingRhombic,
  author  = {Rosen, Ian T. and Painter, Oskar and Houck, Andrew A.},
  title   = {Flat-Band (De)Localization Emulated with a Superconducting Circuit},
  journal = {Physical Review X},
  volume  = {15},
  pages   = {021091},
  year    = {2025},
  doi     = {10.1103/PhysRevX.15.021091}
}

@article{Aidelsburger2013Gauge,
  author  = {Aidelsburger, Monika and Atala, Marcos and Lohse, Michael and Barreiro, Julio T. and Paredes, Bel{\'e}n and Bloch, Immanuel},
  title   = {Realization of the {Hofstadter} Hamiltonian with Ultracold Atoms in Optical Lattices},
  journal = {Physical Review Letters},
  volume  = {111},
  pages   = {185301},
  year    = {2013},
  doi     = {10.1103/PhysRevLett.111.185301}
}

@article{Miyake2013Gauge,
  author  = {Miyake, Hirokazu and Siviloglou, Georgios A. and Kennedy, Colin J. and Burton, William C. and Ketterle, Wolfgang},
  title   = {Realizing the {Harper} Hamiltonian with Laser-Assisted Tunneling in Optical Lattices},
  journal = {Physical Review Letters},
  volume  = {111},
  pages   = {185302},
  year    = {2013},
  doi     = {10.1103/PhysRevLett.111.185302}
}

@article{Dalibard2011,
  author  = {Dalibard, Jean and Gerbier, Fabrice and Juzeli{\=u}nas, Gediminas and {\"O}hberg, Patrik},
  title   = {Colloquium: Artificial gauge potentials for neutral atoms},
  journal = {Reviews of Modern Physics},
  volume  = {83},
  pages   = {1523--1543},
  year    = {2011},
  doi     = {10.1103/RevModPhys.83.1523}
}

@article{Goldman2014,
  author  = {Goldman, Nathan and Dalibard, Jean},
  title   = {Periodically Driven Quantum Systems: Effective Hamiltonians and Engineered Gauge Fields},
  journal = {Physical Review X},
  volume  = {4},
  pages   = {031027},
  year    = {2014},
  doi     = {10.1103/PhysRevX.4.031027}
}

@article{Ozawa2019,
  author  = {Ozawa, Tomoki and Price, Hannah M. and Amo, Alberto and Goldman, Nathan and Hafezi, Mohammad and Lu, Ling and Rechtsman, Mikael C. and Schuster, David and Simon, Jonathan and Zilberberg, Oded and Carusotto, Iacopo},
  title   = {Topological photonics},
  journal = {Reviews of Modern Physics},
  volume  = {91},
  pages   = {015006},
  year    = {2019},
  doi     = {10.1103/RevModPhys.91.015006}
}

@article{Fisher2000BosonLocalization,
  author  = {Fisher, Matthew P. A. and Weichman, Peter B. and Grinstein, Gerald and Fisher, Daniel S.},
  title   = {Boson localization and the superfluid-insulator transition},
  journal = {Physical Review B},
  volume  = {40},
  pages   = {546--570},
  year    = {1989},
  doi     = {10.1103/PhysRevB.40.546}
}

@article{Cartwright2018,
  author  = {Cartwright, Christine and De Chiara, Gabriele and Rizzi, Matteo},
  title   = {Rhombi-chain {Bose--Hubbard} model: Geometric frustration and interactions},
  journal = {Physical Review B},
  volume  = {98},
  pages   = {184508},
  year    = {2018},
  doi     = {10.1103/PhysRevB.98.184508}
}

@article{Koch2023PauliEngine,
  author  = {Koch, Jennifer and Menon, Keerthy and Cuestas, Eloisa
             and Barbosa, Sian and Lutz, Eric and Fogarty, Thom{\'a}s
             and Busch, Thomas and Widera, Artur},
  title   = {A quantum engine in the {BEC--BCS} crossover},
  journal = {Nature},
  volume  = {621},
  pages   = {723--727},
  year    = {2023},
  doi     = {10.1038/s41586-023-06469-8}
}

@article{Chesi2025BosonicEngine,
  author  = {Chesi, Giovanni and Macchiavello, Chiara
             and Sacchi, Massimiliano Federico},
  title   = {Bosonic two-stroke heat engines with polynomial nonlinear coupling},
  journal = {Quantum Science and Technology},
  volume  = {10},
  pages   = {035012},
  year    = {2025},
  doi     = {10.1088/2058-9565/adcd98}
}

@article{Brollo2025PrethermalEngine,
  author  = {Brollo, Alberto and del Campo, Adolfo
             and Bastianello, Alvise},
  title   = {Universal efficiency boost in prethermal quantum heat engines
             at negative temperature},
  journal = {Nature Communications},
  volume  = {16},
  pages   = {10593},
  year    = {2025},
  doi     = {10.1038/s41467-025-66424-1}
}

@article{Uusnakki2026SuperconductingEngine,
  author  = {Uusn{\"a}kki, Tuomas and M{\"o}rstedt, Timm
             and Teixeira, Wallace and Rasola, Miika
             and M{\"o}tt{\"o}nen, Mikko},
  title   = {Initial demonstration of a quantum heat engine based on
             dissipation-engineered superconducting circuits},
  journal = {Nature Communications},
  volume  = {17},
  pages   = {6054},
  year    = {2026},
  doi     = {10.1038/s41467-026-72651-x}
}

@article{Ferreri2025InterferometricEngine,
  author  = {Ferreri, Alessandro and Wang, Hui and Nori, Franco
             and Wilhelm, Frank K. and Bruschi, David Edward},
  title   = {Quantum heat engine based on quantum interferometry:
             The {$SU(1,1)$} {Otto} cycle},
  journal = {Physical Review Research},
  volume  = {7},
  pages   = {013284},
  year    = {2025},
  doi     = {10.1103/PhysRevResearch.7.013284}
}

@article{Curzon1975,
  author  = {Curzon, F. L. and Ahlborn, B.},
  title   = {Efficiency of a Carnot Engine at Maximum Power Output},
  journal = {American Journal of Physics},
  volume  = {43},
  number  = {1},
  pages   = {22--24},
  year    = {1975},
  doi     = {10.1119/1.10023}
}

@article{Hoffmann1997,
  author  = {Hoffmann, K. H. and Burzler, J. M. and Schubert, S.},
  title   = {Endoreversible Thermodynamics},
  journal = {Journal of Non-Equilibrium Thermodynamics},
  volume  = {22},
  number  = {4},
  pages   = {311--355},
  year    = {1997},
  doi     = {10.1515/JNETDY.1997.22.4.311}
}

@article{Wang2017,
  author  = {Wang, Jincheng and He, Jie and He, Xin and Ma, Yonggang},
  title   = {Endoreversible Quantum Heat Engines in the Linear Response Regime},
  journal = {Physical Review E},
  volume  = {96},
  pages   = {012152},
  year    = {2017},
  doi     = {10.1103/PhysRevE.96.012152}
}

@article{Deffner2018,
  author  = {Deffner, Sebastian},
  title   = {Efficiency of Harmonic Quantum Otto Engines at Maximal Power},
  journal = {Entropy},
  volume  = {20},
  number  = {11},
  pages   = {875},
  year    = {2018},
  doi     = {10.3390/e20110875}
}

@article{Smith2020,
  author  = {Smith, Zackary and Pal, Priyo S. and Deffner, Sebastian},
  title   = {Endoreversible Otto Engines at Maximal Power},
  journal = {Journal of Non-Equilibrium Thermodynamics},
  volume  = {45},
  number  = {3},
  pages   = {305--310},
  year    = {2020},
  doi     = {10.1515/jnet-2020-0039}
}

@article{Bouton2021,
  author  = {Bouton, Quentin and Nettersheim, Jan and Adam, David and
             Sch{\"o}nleber, David W. and K{\l}obus, Waldemar and Weimer, Hendrik},
  title   = {An Endoreversible Quantum Heat Engine Driven by Atomic Collisions},
  journal = {Nature Communications},
  volume  = {12},
  pages   = {2063},
  year    = {2021},
  doi     = {10.1038/s41467-021-22222-z}
}

@article{Pena2023,
  author  = {Pe{\~n}a, Francisco J. and Myers, Nathan M. and
             {\'O}rdenes, Daniel and Albarr{\'a}n-Arriagada, Francisco
             and Vargas, Patricio},
  title   = {Enhanced Efficiency at Maximum Power in a Fock--Darwin Model Quantum Dot Engine},
  journal = {Entropy},
  volume  = {25},
  number  = {3},
  pages   = {518},
  year    = {2023},
  doi     = {10.3390/e25030518}
}

@article{Myers2023Graphene,
  author  = {Myers, Nathan M. and Pe{\~n}a, Francisco J. and
             Cort{\'e}s, Natalia and Vargas, Patricio},
  title   = {Multilayer Graphene as an Endoreversible Otto Engine},
  journal = {Nanomaterials},
  volume  = {13},
  number  = {9},
  pages   = {1548},
  year    = {2023},
  doi     = {10.3390/nano13091548}
}

@article{Behrendt2025,
  author  = {Behrendt, Gregory and Deffner, Sebastian},
  title   = {Endoreversible Stirling Cycles: Plasma Engines at Maximal Power},
  journal = {Entropy},
  volume  = {27},
  number  = {8},
  pages   = {807},
  year    = {2025},
  doi     = {10.3390/e27080807}
}

@article{Esposito2010,
  title = {Efficiency at Maximum Power of Low-Dissipation {C}arnot Engines},
  author = {Esposito, Massimiliano and Kawai, Ryoichi and Lindenberg, Katja and Van den Broeck, Christian},
  journal = {Phys. Rev. Lett.},
  volume = {105},
  issue = {15},
  pages = {150603},
  numpages = {4},
  year = {2010},
  month = {Oct},
  publisher = {American Physical Society},
  doi = {10.1103/PhysRevLett.105.150603}
}

@article{Leff1987,
    author = {Leff, Harvey S.},
    title = {Thermal efficiency at maximum work output: New results for old heat engines},
    journal = {American Journal of Physics},
    volume = {55},
    number = {7},
    pages = {602-610},
    year = {1987},
    month = {07},
    issn = {0002-9505},
    doi = {10.1119/1.15071},
    url = {https://doi.org/10.1119/1.15071}
}

\end{document}